\documentclass[pre,
aps,superscriptaddress,floats,showpacs, 10pt]{revtex4-1}

\usepackage{graphicx}
\usepackage{dcolumn}
\usepackage{bm}
\usepackage{amssymb}
\usepackage{amsmath}
\usepackage{color}

\begin{document}

\title{On the breaking of a plasma wave in a thermal plasma: \\ I. The structure of the density singularity}
\author{Sergei V. Bulanov}
\altaffiliation[Also at ]{A. M. Prokhorov Institute of General Physics of RAS, Moscow, Russia}
\affiliation{QuBS, Japan Atomic Energy Agency, 1-8-7 Umemidai, Kizugawa, Kyoto, 619-0215 Japan}
\author{Timur Zh. Esirkepov}
\affiliation{QuBS, Japan Atomic Energy Agency, 1-8-7 Umemidai, Kizugawa, Kyoto, 619-0215 Japan}
\author{Masaki Kando}
\affiliation{QuBS, Japan Atomic Energy Agency, 1-8-7 Umemidai, Kizugawa, Kyoto, 619-0215 Japan}
\author{James K. Koga}
\affiliation{QuBS, Japan Atomic Energy Agency, 1-8-7 Umemidai, Kizugawa, Kyoto, 619-0215 Japan}
\author{Alexander S. Pirozhkov}
\affiliation{QuBS, Japan Atomic Energy Agency, 1-8-7 Umemidai, Kizugawa, Kyoto, 619-0215 Japan}
\author{Tatsufumi Nakamura}
\affiliation{QuBS, Japan Atomic Energy Agency, 1-8-7 Umemidai, Kizugawa, Kyoto, 619-0215 Japan}
\author{Stepan S. Bulanov}
\altaffiliation[Also at ]{Institute of Theoretical and Experimental Physics, Moscow 117218, Russia}
\affiliation{University of California, Berkeley, CA 94720, USA}
\author{Carl B. Schroeder}
\affiliation{Lawrence Berkeley National Laboratory, Berkeley, California 94720, USA}
\author{Eric Esarey}
\affiliation{Lawrence Berkeley National Laboratory, Berkeley, California 94720, USA}
\author{Francesco Califano}
\affiliation{Physical Department, University of Pisa, Pisa 56127, Italy}
\author{Francesco Pegoraro}
\affiliation{Physical Department, University of Pisa, Pisa 56127, Italy}
\date{19/Apr/2012, 12:30, Japan time}

\begin{abstract}
The structure of the  singularity that is formed in a  relativistically large amplitude plasma wave  close 
to the wavebreaking limit is found by 
 using a simple waterbag electron distribution function. 
 The electron density distribution
in the  breaking wave has a typical ``peakon" form. 
The maximum value of the electric field  in a thermal breaking plasma 
is obtained and compared to the  cold plasma limit.  
The results of computer simulations for different initial 
electron distribution functions are in agreement with the theoretical conclusions.
\end{abstract}
\maketitle

\section{Introduction}

Finite amplitude waves in a plasma have been studied intensively for decades
in regard to a broad range of physical problems related to astrophysics,
magnetic  and inertial confinement thermonuclear fusion and in nonlinear
wave theory \cite{GINZBURG}. In particular, nonlinear plasma waves are of crucial importance for wakefield
acceleration  in plasma configurations where the wakewave is generated either by laser pulses \cite{LWFA,ESL1} 
or  by bunches of relativistic electrons \cite{PWFA}, 
for high-harmonic generation \cite{HOH}  and for many other aspects of
 laser-plasma physics \cite{MTB, Part-II}. In order to support a  strong electric
field the Langmuir wave must be highly nonlinear. In  a stationary wave the
limit on  the field  amplitude is imposed by the wave breaking condition \cite{AP}, while in 
a nonstationary wave in the regime beyond the wavebreaking point the electric field 
can be even higher \cite{MaX}.

Nonlinear wave breaking  exhibits  one of the fundamental phenomena in
the mechanics of continuous media. When the wave  amplitude approaches  and/or exceeds the breaking limit   
the wave form becomes singular as its profile steepens,
finally leading to the formation of a multi-stream motion. Even in the  simplest case of
one-dimensional electrostatic Langmuir waves in collisionless plasmas this
process still attracts  great interest  due to its importance both  for the wave amplitude 
limitation \cite{AP, RCD, JMD, KM} and  for  its practical relevance to the electron injection
into the wakefield acceleration phase \cite{MaX, Inj}. In the
application to the laser wake field acceleration   attention is paid mainly 
to the determination of   the upper limit for the electric field \cite{KM, MaX, ESL2, TR,
COF, BurNob, SMF}.

Thermal effects in a warm plasma can reduce the maximum wave amplitude 
\cite{KM, MaX, ESL2, TR, COF, BurNob, SMF} and modify the character of the  singularity \cite{SMF}. A  finite plasma
temperature limits the electron density in the breaking wave but in the  general
case does not necessarily  lead to smooth density distributions. 
Since the results obtained by B. Riemann in  the 19$^{th}$
century on the  wave breaking of nonlinear sound waves  (see Ref. \cite{BRMNN}, 
and \cite{LLHd}), it has been known that  thermal effects do not
prevent the ``gradient catastrophe". In this case the singularity
in the breaking wave corresponds to  a  shock-like wave profile. Other remarkable
singularities are known for nonlinear waves on a water surface which at the breaking points
become of the type of 
``Stokes's traveling crested extreme wave'' with the interior crest angle of $2 \pi/3$ \cite{Stokes, Whitham}.
We also note here  the exact solutions,  known as ``peakons'', of nonlinear partial differential equations
 describing the waves on shallow water that  have the  form of 
a soliton with  a discontinuous first derivative \cite{peakon}.

In the present paper  we analyze  the
structure of the Langmuir wave breaking  and  show that  crested Langmuir waves in thermal plasmas 
have a profile with  a discontinuous first derivative.

\section{The water-bag model for a relativistic Langmuir wave in a thermal plasma}

\subsection{Electron distribution function formed as a result of a gas multiphoton ionization }

\begin{figure}[tbph]
\includegraphics[width=6cm,height=4cm]{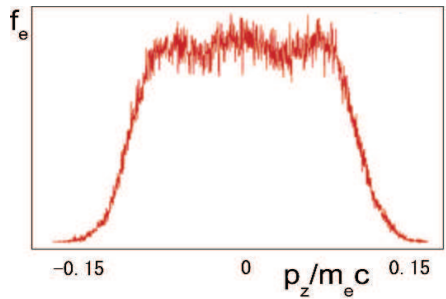}
\caption{Electron distribution function $f_e(p_z)$ (arb. units)
formed as a result of optical field ionization of the gas target by a $40\, fs$  
laser pulse with dimensionless
amplitude $eE/m_e\omega_0c=0.1$, which is the tunneling regime of $\gamma_K\ll 1$ \cite{KCP}. 
The $f_e$ dependence is on the normalised 
electron momentum $p_z/m_ec$ along the laser polarization direction.}
\label{fig1}
\end{figure}

 In the case of  a plasma irradiated by  a high intensity laser  pulse the temperature  is 
determined by the laser light parameters for the time interval before the main pulse comes.
During the  interaction  of a  femtosecond, terawatt laser pulse with gas targets  a plasma is created via 
photoionization \cite{KCP} by the prepulse or by  the ASE (Amplified Spontaneous Emission) pedestal. 
In such a
collisionless plasma the electron energy is of the order of the quiver energy
in the ionizing laser field, i.e. typically in the range below $keV$. It is thus 
substantially lower than the electron energy in the main laser pulse, which is typically in the $MeV$ range,
that excites  the wake plasma wave (e.g.,  see Fig. \ref{fig1}, where a typical electron 
distribution function formed as a result of optical field ionization of the gas target 
by an ultrashort laser pulse is shown, \cite{Koga2011}). 
Being limited by the quiver energy,  the electron 
distribution function is not Maxwellian and can be adequately described by a
simple water-bag model, which is otherwise considered to be too
artificial and restrictive. We note that the water-bag electron distribution
function has been used in Refs. \cite{KM, MaX, TR, COF, BurNob}.

\subsection{Basic equations}
 
\begin{figure}[tbph]
\includegraphics[width=8cm,height=5.3cm]{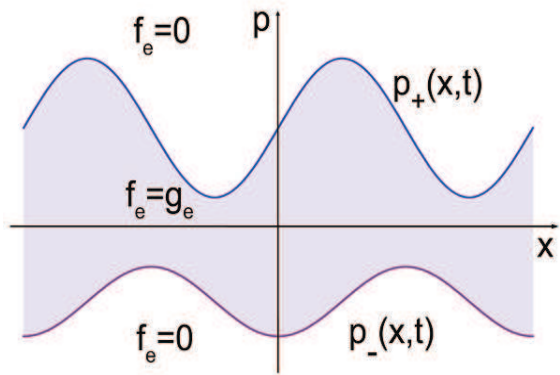}
\caption{Electron phase space in a single water-bag model. 
The electron distribution function is constant $f_{e}(p,x,t)=g_e$
within the region with borders marked by $p_{+}(x,t)$ and $p_{-}(x,t)$, 
while  $f_{e}(p,x,t)=0$ outside this region.}
\label{fig2}
\end{figure}

Following Ref. \cite{RCD} 
, we consider the electron phase space $(x,p)$ shown 
in Fig. \ref{fig2}, which corresponds to the support of the electron distribution function $f_e(p,x,t)$.

The electron distribution function is constant 
\begin{equation}
f_{e}(p,x,t)={ \rm constant}  \label{eq0-fe}
\end{equation}
within the region with borders marked by $p_{+}(x,t)$ and $p_{-}(x,t)$, 
while  $f_{e}(p,x,t)=0$ outside this region. Here the constant is
proportional to the ratio of the electron density and the momentum width.
The electron distribution function can also be expressed via the unit step Heaviside functions: 
\begin{equation}
f_{e}(p,x,t)=
{\rm constant}  
\times  \theta(p-p_-(x,t))\, \theta(p_+(x,t)-p), 
\end{equation}
where $\theta(x)=0$ for $x<1$ and $\theta(x)=1$ for  $x>1$.

The evolution  of the distribution function is described
by the Vlasov-Poisson system of equations%
\begin{equation}
\partial _{t}f_{e}+v\partial _{x}f_{e}-E\, \partial _{p}f_{e}=0,  \label{eq1-Vl}
\end{equation}
\begin{equation}
\partial _{x}E=1-n_{e},  \label{eq2-Poi}
\end{equation}
where all the variables are written in  a dimensionless form normalised in
a standard way  in which the time and space units are $\omega
_{pe}^{-1}$ and $c\omega _{pe}^{-1}$, the momentum and the velocity are
normalised on $m_{e}c$ and $c$, the unit for the electric field, $E(x,t)$, 
is $m_{e}\omega _{pe}c/e$, with $\omega _{pe}=(4\pi
n_{0}e^{2}/m_{e})^{1/2}$ being the Langmuir frequency, $e$ and $m_{e}$ are the electron
charge and mass, and $n_{0}$ is the density of
ions which are assumed to be at rest. The electron
velocity is equal to $v=p/(1+p^{2})^{1/2}$ , and $n_{e}(x,t)$ is the electron
density normalised on $n_{0}$. Global charge neutrality is assumed. 
Eq. (\ref{eq1-Vl}) describes the
incompressible motion of the distribution $f_{e}$ in phase space.

Calculating the first momentum of the distribution function we  find that the
 electron density is related to the bounding curves  $p_{+}(x,t)$ 
and $p_{-}(x,t)$ as 
\begin{equation}
n_{e}(x,t)=\int\limits_{-\infty }^{+\infty }f_{e}(p,x,t) d{\rm p} =g_e\, [p_{+}(x,t)-p_{-}(x,t)],
\label{eq3-ne}
\end{equation}%
where  $g_e$ is   a  numerical constant (see Eq.(\ref{eq0-fe}))
that gives  the ratio between   the  dimensionless electron density $n_e(x,t)$ 
and the  dimensionless  momentum width   $\Delta p(x,t) \equiv  p_{+}(x,t)- p_{-}(x,t)$,
and is determined by the value of this ratio at $t=0$.

From Eqs. (\ref{eq1-Vl}) and (\ref{eq2-Poi}) it follows that 
the functions  $p_{+}(x,t)$, $p_{-}(x,t)$ and $E(x,t)$ evolve according to (see also Ref.\cite{RCD}) 
\begin{equation}
\partial _{t}p_{+}+\frac{p_{+}}{\sqrt{1+p_{+}^{2}}}\partial _{x}p_{+}=-E,
\label{eq4-ppl}
\end{equation}%
\begin{equation}
\partial _{t}p_{-}+\frac{p_{-}}{\sqrt{1+p_{-}^{2}}}\partial _{x}p_{-}=-E,
\label{eq5-pmns}
\end{equation}%

\begin{equation}
\partial _{x}E=1-g_e(p_{+}-p_{-}).
\label{eq6-E}
\end{equation}

 \subsection{Dispersion equation for the wave frequency and wave number}
 
A large energy spread of the electron
distribution function leads to  a change
of the Langmuir wave frequency due to its dependence on the plasma temperature. 
Linearization of Eqs. (\ref{eq4-ppl} -- \ref{eq6-E}) around  
the equilibrium solution $p_{+}=p_{+,0}$, $p_{-}=p_{-,0}$, $E=0$
with $p_{-,0}=- p_{+,0}=\Delta p_0/2$ gives the dispersion equation for the frequency, $\omega$, 
and  the wave-number $k$ in the case of the small amplitude Langmuir wave.  
In  dimensional form it can be written  as
\begin{equation}
\omega^2=\frac{8\pi n_0 e^2 c}{\sqrt{4 m_e^2c^2+\Delta p_0^2}}+ \frac{k^2 c^2 \Delta p_0^2}{ 4 m_e^2c^2+\Delta p_0^2}.
\label{eq25-omthrml}
\end{equation}
The corresponding kinetic dispersion relation for a relativistic Maxwellian distribution function
 (J\"uttner-Synge distribution) and its derivation in terms of relativistic fluid-like equations are 
 given in Ref. \cite{PORPEG} and references quoted therein.  In particular waves with phase 
 velocities larger than the speed of light are considered in Ref. \cite{PORPEG} since in the case 
 of the J\"uttner-Synge distribution, and in general of a  distribution that is not piece-wise 
 constant in momentum space, waves with phase velocity smaller that the speed of light are heavily 
 damped in the relativistic regime \cite{buti}. It is shown that for these ``superluminal''  waves the
relativistic electron population obeys an effective isothermal equation of state as soon as the 
normalized thermal momentum $p_{th}$ becomes larger than an appropriately redefined phase
momentum $p_{ph}\equiv \omega/(\omega^2 - k^2 c^2)^{1/2}$. We note that although the superluminal
regime could also be considered within the water-bag formalism by  redefining $\gamma_{ph}$ 
below Eq.(\ref{eq10-pmh}),  such a  regime is not of interest for the investigation of wavebreaking. 
In fact,  both Landau damping and wavebreaking  are related to particles that have  an unperturbed velocity 
(in the case of the linear Landau damping), or are accelerated  by the wave  electric field to  
a velocity that matches the wave phase velocity and this  cannot occur for superluminal waves. 
Conversely, this indicates that thermal effects tend to favour Landau damping in the case 
of  a particle distribution that is not piece-wise constant  and  wavebreaking in the case 
of a water-bag distribution. 
    
	As can be  seen from Eq.(\ref{eq25-omthrml}) a finite temperature modifies the Langmuir frequency and 
makes   it  depend  on the wavenumber, $k$.
In terms of the variable $X=x-v_{ph}t$  the wave is characterised by the wavenumber $k_w$, 
which is given in dimensionless form by
\begin{equation}
k_w=\sqrt{\frac{\gamma_{+,0}}{\beta_{\rm ph}^2 \gamma_{+,0}^2-p_{+,0}^2}},
\label{eq25-omthrmlkw}
\end{equation}
where $\gamma_{+,0}=\sqrt{1+p_0^2}$. The wave number $k_w$ tends to infinity for $p_{+,0}/\gamma_{+,0} \to \beta_{\rm ph}$ .
The frequency dependence on the  electron temperature leads
to a shortening of the wakewave wavelength.
It results in a lower electric field in the wakewave  
in comparison  to the case with a relatively small thermal spread.

\subsection{Long wavelength limit}

It is easy 
to obtain from Eqs. (\ref{eq4-ppl} -- \ref{eq6-E}) 
that spatially homogeneous nonlinear oscillations of electrons in relativistic 
thermal plasmas
are described by the system of ordinary differential equations 
\begin{equation}
\frac{dp_{+}}{dt}=-E,
\end{equation}
\begin{equation}
\frac{dE}{dt}=\frac{\sqrt{1+p_{+}^2}-\sqrt{1+(p_{+}-\Delta p_0)^2}}{\Delta p_0},
\end{equation}
where $\Delta p_0=p_{+,0}-p_{-,0}$. The Hamilton function  corresponding to these equations  is
\begin{equation}
{\cal H}(E,p_{+})=\frac{E^2}{2}+{\Pi}(p_{+})
\label{eq26-HamT0}
\end{equation}
with the potential function
\begin{equation}
{\Pi}(p_{+})=\frac{U(p_{+})-U(p_{+}-\Delta p_0)}{2\Delta p_0},
\label{eq26-PotT0}
\end{equation}
where the function $U(z)$ is given by 
\begin{equation}
U(z)=z\sqrt{1+z^2}+\ln{\left(z+\sqrt{1+z^2}\right)}.
\end{equation}

\begin{figure}[tbph]
\includegraphics[width=14cm,height=4cm]{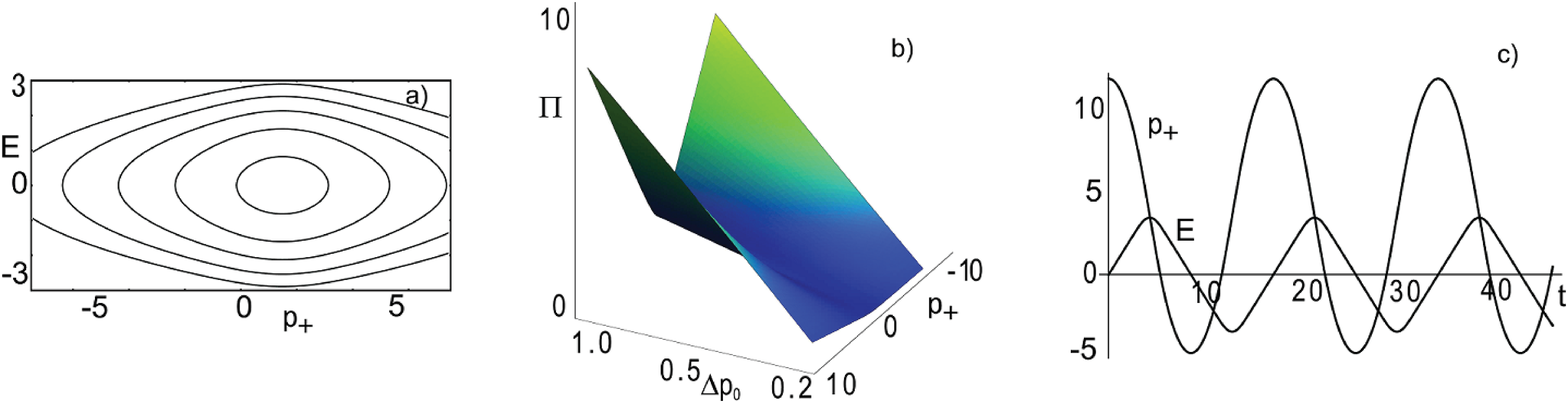}
\caption{a) Isocontours of the Hamiltonian function (\ref{eq26-HamT0}) in the plane $(E,p_{+})$ for $\Delta p_0=3$;
b) Potential function ${\Pi}(p_{+};\Delta p_0)$;
c) Time dependence of the electron momentum, $p_{+}(t)$ and electric field, $E(t)$, 
for $p_{+,0}=12.5$ and $\Delta p_0=7.5$.}
\label{fig3}
\end{figure}

Isocontours of the Hamiltonian function (\ref{eq26-HamT0}) 
in the plane $E,p_{+}$ are shown in Fig. \ref{fig3}a for $\Delta p_0=3$.
The potential function ${\Pi}(p_{+})$ is plotted in Fig. \ref{fig3}b.
Nonlinear oscillations are shown in Fig. \ref{fig3}c, 
where  the time dependence of the electron momentum, $p_{+}(t)$ 
and electric field, $E(t)$ are plotted for $p_{+,0}=12.5$ and $\Delta p_0=7.5$.
The momentum $p_+$  oscillates between the value $p_{+,0}$ and $-(p_{+,0}-\Delta p_0)$.

In the small amplitude limit the oscillation frequency is given by expression 
(\ref{eq25-omthrml}) for $k=0$, i.e. 
$\omega=\frac{\left({8\pi n_0 e^2 c}\right)^{1/2}}{{\left(4 m_e^2c^2+\Delta p_0^2 \right)^{1/4}}}$.
If the oscillation amplitude, $p_{\rm m}$, is large compared  to the half-width of the 
distribution function, $\Delta p_0/2$,
and $p_{\rm m} \gg 1$ then the oscillation period 
is given by $T=(4 p_{\rm m}/\pi n_0 e^2)^{1/2}$ (see Ref. \cite{AP}). 

 \subsection{Langmuir waves travelling with constant velocity}

We consider waves propagating along the $x$ axis with constant phase
velocity $v_{ph}$, where  all functions depend on the independent 
variable 
\begin{equation}
X=x-\beta_{\rm ph}t.
\label{eq7-X}
\end{equation}%
In this case Eqs. (\ref{eq4-ppl} -- \ref{eq6-E}) take the form
\begin{equation}
h_{+}'=-E,
\label{eq8-hpl}
\end{equation}%
\begin{equation}
h_{-}'=-E,
\label{eq8-hm}
\end{equation}%
and
\begin{equation}
E'=1-\frac{p_{+}(h_{+})-p_{-}(h_{-})}{\Delta p_0},
\label{eq9-E}
\end{equation}%
where we introduced the dependent variables  $h_+$ and $h_-$ defined by
\begin{equation}
h_{\pm}(p_{\pm})=\sqrt{1+p_{\pm}^{2}}-\beta_{\rm ph}p_{\pm}, 
\label{eq10-hpm}
\end{equation}
a "prime" denotes differentiation with respect to $X$ and $\beta_{\rm ph}=v_{ph}/c$.
We use 
\begin{equation}
\Delta p_0=p_{+,0}-p_{-,0},
\end{equation}
 with $p_{+,0}=p_{+}(X_0)$ 
and $p_{-,0}=p_{-}(X_0)$ taken at $X=X_0$,   where $E'=0$. 
Inverting  Eq. (\ref{eq10-hpm})  we obtain
\begin{equation}
p_{\pm}(h_{\pm})=\gamma_{\rm ph}^2\beta_{\rm ph}h_{\pm}-\sqrt{\gamma_{\rm ph}^4 h_{\pm}^2-\gamma_{\rm ph}^2},
\label{eq10-pmh}
\end{equation}
with $\gamma_{\rm ph}=1/\sqrt{1-\beta_{\rm ph}^2}$.

Here and below we assume "subluminal" propagation velocity, i.e. $\beta_{\rm ph} \leq 1$.

\subsection{Hamiltonian form of  the equations describing a  travelling Langmuir wave}

Multiplying Eq. (\ref{eq9-E}) by  $E$ and using Eqs. (\ref{eq8-hpl}) and (\ref{eq8-hm}) 
and integrating it over  $X$, we obtain the integral
\begin{equation}
\displaystyle{\frac{E^2}{2} +\frac{h_{+}}{2}\left(1-\frac{\gamma_{\rm ph}^2\beta_{\rm ph}h_{+}}{\Delta p_0} \right)
+\frac{h_{-}}{2}\left(1+\frac{\gamma_{\rm ph}^2\beta_{\rm ph}h_{-}}{\Delta p_0} \right)
}
+\displaystyle{\frac{W(\gamma_{\rm ph} h_{+})-W(\gamma_{\rm ph} h_{-})}{2\Delta p_0}={\rm constant}},
\label{eq10-intb}
\end{equation}
where  
\begin{equation}
W(z)=z\sqrt{z^2-1}\, -\, \ln{\left(z+\sqrt{z^2-1}\right)}.
\label{eq10-Wb}
\end{equation}
This function vanishes at $z=1$. In the limit $z\to 1+0$ its behaviour is described as 
\begin{equation}
W(z)=\frac{4}{3}\sqrt{2}(z-1)^{3/2} +O(z-1)^{5/2}.
\label{eq10-Wb0}
\end{equation}
For $z \to \infty$ we have
\begin{equation}
W(z)=z^2-\frac{1}{2}-\ln{2 z}+\frac{1}{8 z^2} +O\left(\frac{1}{z^3}\right).
\label{eq10-Wbinf}
\end{equation}

According to Eqs. (\ref{eq8-hpl}) and (\ref{eq8-hm}) the variables  $h_{-}$ and $h_{+}$, 
are not independent and  are related by
\begin{equation}
h_{-}=h_{+}-\Delta h_0,
\end{equation}
where the constant $\Delta h_0$ is determined by  the values of $p_+$ and $p_-$ at $X=X_0$:
\begin{equation}
\displaystyle{\Delta h_0=h_{+,0}-h_{-,0}}
\displaystyle{\equiv \sqrt{1+p_{+,0}^{2}}-\sqrt{1+p_{-,0}^{2}}-\beta_{\rm ph}(p_{+,0}-p_{-,0})}.
\label{h0}
\end{equation}
As a result,  Eqs. (\ref{eq8-hpl}, \ref{eq8-hm}) and (\ref{eq9-E}) can be rewritten in the form 
\begin{equation}
h_{+}'=-E,
\label{eq12-hppr}
\end{equation}%
\begin{equation}
E'=1-
\frac{p_{+}(h_{+})-p_{-}(h_{+}-\Delta h_0)}{\Delta p_0}.
\label{eq13-Eh}
\end{equation}%
This is a  Hamiltonian system with Hamilton function
\begin{equation}
{\cal H}(E,h_{+})=
\displaystyle{\frac{E^2}{2}+\Pi (h_{+})},
\label{eq14-Ham}
\end{equation}%
where $h_{+}$ and $-E$ are canonical variables, and 
\begin{equation}
\Pi (h_{+})=
\displaystyle{h_{+} \left( 1-\frac{\gamma_{\rm ph}^2 \beta_{\rm ph}\Delta h_0}{\Delta p_0} \right)}
\displaystyle{+\, \frac{W(\gamma_{\rm ph} h_{+})-W(\gamma_{\rm ph} (h_{+}-\Delta h_0))}{2\Delta p_0}}.
\label{eq14-POT}
\end{equation}

For a  symmetrical  distribution  where $p_{+,0}= -p_{-,0}$ Eq. (\ref{h0}) 
takes the simpler form $ \Delta h_0=-\beta_{\rm ph}\Delta p_{0}$ 
and the potential $ \Pi (h_{+})$ reduces to
\begin{equation}
\Pi_{\rm sym} (h_{+})=
\displaystyle{\gamma_{\rm ph}^2  h_{+} } 
\displaystyle{+\, \frac{W(\gamma_{\rm ph} h_{+})-W(\gamma_{\rm ph} (h_{+} + \beta_{\rm ph} \Delta p_0))}{2\Delta p_0}}.
\label{eq14-POTsym}
\end{equation}

\begin{figure}[tbph]
\includegraphics[width=10.4cm,height=4cm]{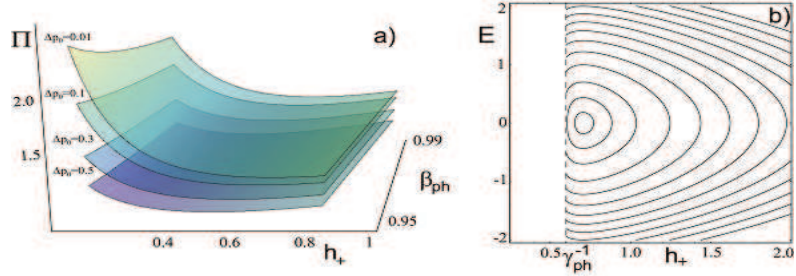}
\caption{a) {Potential function $\Pi(h_{+})$ vs $h_{+}$ and $\beta_{\rm ph}$ for $\Delta p_0=0.01, 0.1, 0.3, 0.5$;
 b) Isocontours of the Hamiltonian function in the plane $(E,h_{+})$ for $\beta_{\rm ph}=0.8$ 
 and $\Delta p_0=0.1$}. }
\label{fig4}
\end{figure}

The potential  $\Pi(h_{+})$ is plotted in Fig. \ref{fig4}a as a function 
of $h_{+}$, $\beta_{\rm ph}$ for four
values of $\Delta p_0$ assuming that $p_{+,0}= - p_{-,0}$, i.e. $\Delta h_0=- \beta_{\rm ph} \Delta p_0$.
Isocontours of the Hamiltonian function in the plane $E,h_{+}$ for $\beta_{\rm ph}=0.8$ and $\Delta p_0=0.1$ 
are shown in Fig. \ref{fig4}b.

\section{The wavebreaking limits}

\subsection{Crested Langmuir wave}
\label{cres}

The system of Eqs. (\ref{eq8-hpl}-\ref{eq9-E}) has a singular solution when $h_{+} \to \gamma_{\rm ph}^{-1}$, 
i.e.
\begin{equation}
p_{+}\to p_{+,{\rm br}}=\frac{\beta_{\rm ph}}{\sqrt{1-\beta_{\rm ph}^2}},
\label{eq13-pbr}
\end{equation}%
which corresponds to the wavebreak in thermal plasmas when the electron 
velocity calculated for the momentum on the upper bound curve,
$p_+(x,t)$, becomes equal to the wave phase velocity.
In this  limit  $d h _+/d p_+ \to 0$ and  the upper bound curve is no longer  
a  single valued function of $X$.

The electron momentum on the lower bound curve at wavebreak  is 
\begin{equation}
p_{-,br}=p_{+,{\rm br}}-\beta_{\rm ph}\gamma_{\rm ph}^2\Delta h_0-\sqrt{\gamma_{\rm ph}^4 \Delta h_0^2-2\gamma_{\rm ph}^3\Delta h_0}.
\label{eq14-pmpbr}
\end{equation}%

Eqs. (\ref{eq3-ne}, \ref{eq13-pbr}, \ref{eq14-pmpbr}) give for the electron density
at the wavebreaking point
\begin{equation}
n_{e,{\rm br}}=\displaystyle{\frac{\sqrt{\gamma_{\rm ph}^4 \Delta h_0^2-2\gamma_{\rm ph}^3\Delta h_0}
+\beta_{\rm ph}\gamma_{\rm ph}^2 \Delta h_0}{\Delta p_0}}.
\label{eq15-nebr}
\end{equation}%

For a symmetric distribution function such that $p_{+,0}= - p_{-,0}$   
from the electron density dependence on $h$,
\begin{equation}
n_{e}(h)
=
\frac{\gamma_{\rm ph} \displaystyle{\left[\sqrt{\gamma_{\rm ph}^2(h+\beta_{\rm ph}\Delta p_0)^2-1}-
\sqrt{\gamma_{\rm ph}^2h^2-1}\right]}}{\Delta p_0}-\gamma_{\rm ph}^2\beta_{\rm ph}^2,
\label{eq15-nevsh}
\end{equation} 
it follows that at the wavebreaking point, $h\to \gamma_{\rm ph}^{-1}$, the density tends to 
\begin{equation}
n_{e,{\rm br}}=
\gamma_{\rm ph}^2 \beta_{\rm ph}
\left(\sqrt{1+\frac{2}{\beta_{\rm ph}\gamma_{\rm ph}\Delta p_0}}-\beta_{\rm ph}\right).
\end{equation} 

In the nonrelativistic limit,  when $\beta_{\rm ph}\ll 1$, $\Delta p_0 \ll 1/\beta_{\rm ph}$, 
and $\gamma_{\rm ph}\approx 1$, the density is
\begin{equation} 
n_{e,{\rm br}}\approx \sqrt{\frac{2\beta_{\rm ph}}{\Delta p_0}} -\gamma_{\rm ph}^2\beta_{\rm ph}^2.
\label{eq16-nebrcl}
\end{equation}%

In the ultrarelativistic limit, when $\beta_{\rm ph}\approx 1$, and $\gamma_{\rm ph} \gg 1$ we have 
\begin{equation}
n_{e,{\rm br}}\approx \sqrt{\frac{2 \beta_{\rm ph}\gamma_{\rm ph}^3}{\Delta p_{0}}} -\gamma_{\rm ph}^2\beta_{\rm ph}^2
\label{eq17-nebrrl}
\end{equation}%
provided 
$\Delta p_0 \ll  {2}/ \beta_{\rm ph}\gamma_{\rm ph}$  (see also \cite{SMF})
while for $\Delta p_0 =  2 \beta_{\rm ph}\gamma_{\rm ph}$ we have $n_{e,{\rm br}}=1$,
because in this limit a wave with arbitrarily small amplitude breaks, as seen from Eq. (\ref{eq25-omthrmlkw}). 
In the above considered cases the electron density written in dimensional units is 
\begin{equation}
n_{e,{\rm br}}\approx n_0 \sqrt{\frac{ m_e v_{\rm ph}}{p_{+,0}}} 
\label{eq17a-nbr}
\end{equation}%
for a nonrelativistic plasma wave and 
\begin{equation}
n_{e,{\rm br}}\approx n_0 \sqrt{\frac{m_e c \beta_{\rm ph} \gamma_{\rm ph}^3}{p_{+,0}}}
\label{eq17b-nbr}
\end{equation}%
 in the limit $\gamma_{\rm ph} \gg 1$.
 
\begin{figure}[tbph]
\includegraphics[width=6cm,height=10cm]{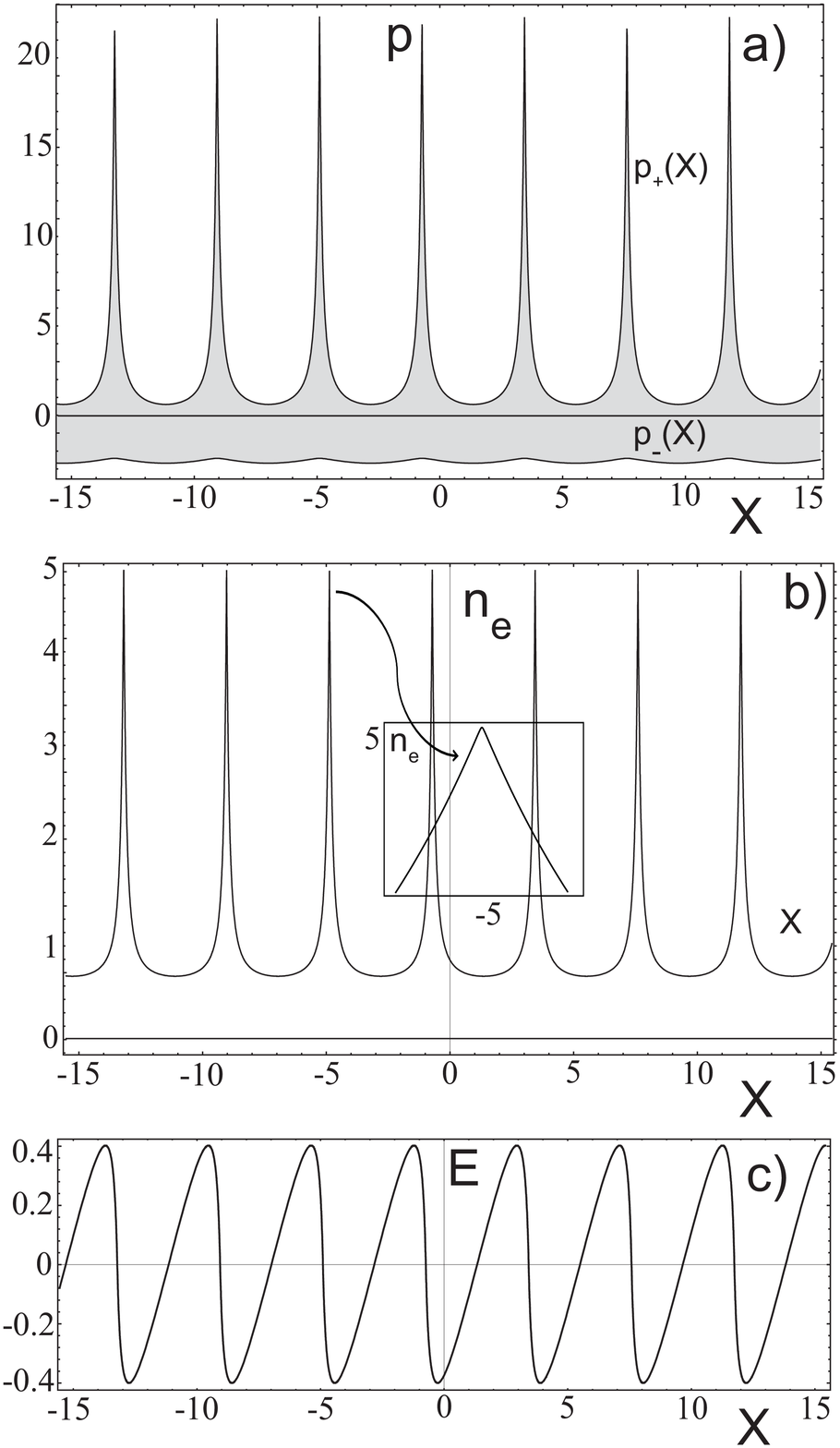}
\caption{Structure of the nonlinear wake wave: 
a) electron phase space, 
b) the electron density 
(in the inset the density distribution is shown in the vicinity of the maximum), 
c) electric field
as functions of the coordinate $X$. The normalised wave phase velocity is $\beta_{\rm ph}=0.999$;
the plasma thermal momentum width at $X=15$ is $\Delta p_0=p_{+,0}-p_{-,0}=5$;
the maximum of the electric field is $E_{\rm m}=0.4190625$.}
\label{fig5}
\end{figure}

In order to find the density behaviour in the neighbourhood of 
the breaking point, we expand the electron momentum, $p_{+}$, on the upper bound curve, 
in the vicinity of its maximum, $\delta X=X-X_{\rm br} \to 0$. Here $X_{\rm br}$ 
is the location of  the breaking point. 
Locally, the momentum is represented by 
\begin{equation}
p_{+}=p_{+,{\rm br}}-\delta p_{+} + O(\delta p_{+}^2)
\label{eq18-pmpbr}
\end{equation}%
with $p_{+,{\rm br}}$ given by Eq. (\ref{eq13-pbr}).
Keeping the main terms of  the expansion over $\delta p_{+}$ of Eq. (\ref{eq15-nevsh}) 
we obtain for the electron density
\begin{equation}
n_e=n_{e,{\rm br}} -\frac{\delta p_{+}}{\Delta p_0},
\label{eq19-npsing}
\end{equation}
where we used the expression $h \approx  \gamma_{\rm ph}^{-1} +\delta p_{+}^2/2\gamma_{\rm ph}^{3}$.
From Eqs. (\ref{eq12-hppr}, \ref{eq13-Eh}) for the dependence of $\delta p_{+}$ on $X$ we have 
\begin{equation}
(\delta p_{+}^2)^{\prime \prime}=2n_{e,{\rm br}} \gamma_{\rm ph}^3.
\label{eq19-psing}
\end{equation}%
Integrating this expression we find
\begin{equation}
\delta p_{+}=\pm \sqrt{n_{e,{\rm br}} \gamma_{\rm ph}^3} \delta X,
\label{eq20-psinX}
\end{equation}%
where we assumed that $\delta p_{+}^{\prime}$ at $\delta X$ vanishes, i.e. the electric field 
at the breaking point is equal to zero.
Since by assumption $\delta p_{+}$ must be non-negative,
we  must  chose the $"-"$  sign in the interval $\delta X<0$ and the $"+"$  sign for $\delta X>0$.
As a result we can write for the momentum $p_{+}$ in the vicinity of the wavebreaking point
\begin{equation}
p_{+}\sim p_{+,{\rm br}}-\sqrt{n_{e,{\rm br}} \gamma_{\rm ph}^3} |\delta X|.
\label{eq21-psinX}
\end{equation}%
From the expression (\ref{eq3-ne}) for the density, and   recalling that 
at wavebreak $\delta p_- \propto (\delta p_+)^2 $, 
we find that  in the vicinity of the breaking point  the electron density can be written as (see also \cite{SMF})
\begin{equation}
n_{e}\sim  {n_{e,{\rm br}}} - \frac  {\sqrt{n_{e,{\rm br}} \gamma_{\rm ph}^3} } {\Delta p_0}\,  |\delta X| .
\label{eq23-nebr}
\end{equation}%

This type of  wave breaking in the  general case corresponds to the "peakon" 
structures known in water waves
\cite{Stokes, Whitham, peakon}. It can also be called "$\Lambda$-type" breaking.

The structure of the nonlinear wake wave of  the electron density, 
the electron phase space and the electric field
is shown  in Fig. \ref{fig5}, as  obtained by numerical integration of 
Eqs. (\ref{eq4-ppl} -- \ref{eq6-E}).
In Fig. \ref{fig5} we show the high temperature case with the initial distribution 
function width $p_{+,0}$ comparable with 
the value of electron momentum on the upper bound curve at the wavebreaking point, $p_{\rm br}$.
From Fig. \ref{fig5}b 
we see that in the vicinity of the density 
maximum (see inset to Fig. \ref{fig5}b), 
the density  dependence on $X$ corresponds to Eq. (\ref{eq23-nebr}).

\subsection{Maximum electric field in stationary wave}

\begin{figure}[tbph]
\includegraphics[width=9.6 cm, height=4.2 cm]{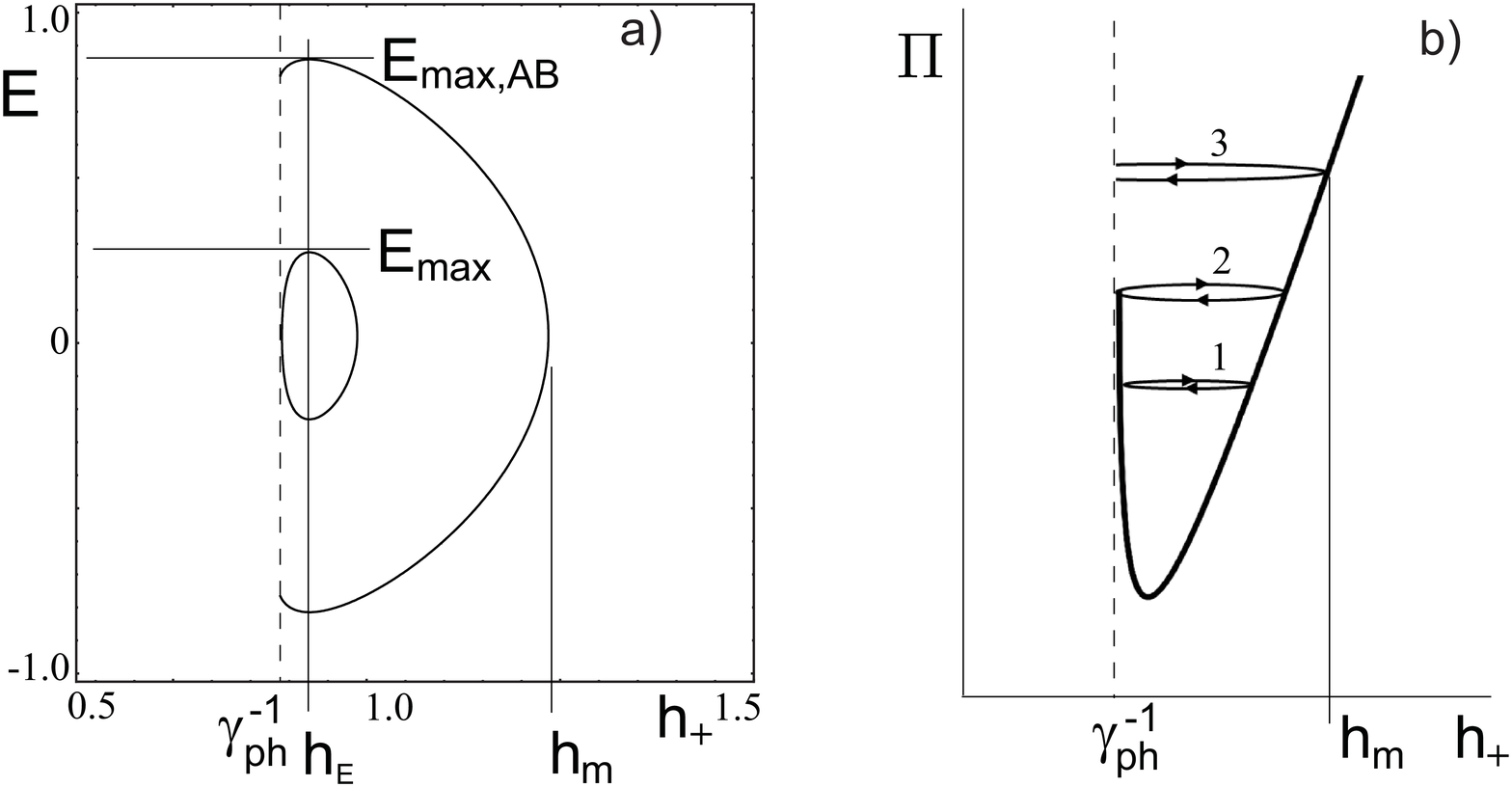}
\caption{ a) Isocontours of the Hamiltonian function ($H_1=1.9125$ and $2.25$) in the plane $(E,h_{+})$ 
for $\beta_{\rm ph}=0.6$ and $\Delta p_0=0.05$. The electric field maxima 
$E_{\max}$ and $E_{\rm max, AB}$ are reached at $h_{+}=h_E$.
b) The curves $1,2,3$ indicate the potential $\Pi(h)$ values  corresponding to 
(1) a periodic wave  with  amplitude below
the wave breaking limit, (2) to a  wave at the wave breaking threshold, and 
(3) to a wave  with  amplitude
above the wave breaking limit. Vertical dashed line marks the wave breaking boundary.  
}
\label{fig6}
\end{figure}

As  is seen from the trajectory pattern in the $E,h_{+}$ plane presented in Fig. \ref{fig4}b, 
the electric field maximum is reached at the point $h_{+}=h_{E}$ (see also Fig. \ref{fig6}) where
the derivative of the electric field with respect to $h_{+}$ vanishes, 
$\left.dE/dh_{+}\right\vert_{h_{+}=h_{E}}=0$. This condition results in the equation for $h_{E}$:
\begin{equation}
\Delta p_0=p_{+}(h_{E})-p_{+}(h_{E}+\beta_{\rm ph}\Delta p_0),
\label{eqhE0}
\end{equation}
where the function $p_{+}(h_{E})$ is given by Eq. (\ref{eq10-pmh}). 
Here we assume the symmetric distribution 
with $p_{+,0}=-p_{-,0}=\Delta p_0/2$. The solution of Eq. (\ref{eqhE0}) is
\begin{equation}
h_{E}=\sqrt{1+\frac{\Delta p_0^2}{4}}-\frac{\beta_{\rm ph}\Delta p_0}{2}.
\label{eqhE}
\end{equation}

The last bound trajectory in the $E,h_{+}$ plane is determined by the equation
\begin{equation}
\frac{E^2}{2}+\Pi(h_{+})=\Pi(\gamma_{\rm ph}^{-1})
\label{eqlast}
\end{equation}
with the potential given by Eq. (\ref{eq14-POTsym}).
Substituting $h_{+}=h_{E}$ we find the electric field maximum
\begin{equation}
E_{\max}=\sqrt{2\left[\Pi(\gamma_{\rm ph}^{-1})-\Pi(h_{E})\right]}.
\label{eqEmax}
\end{equation}

In the limit of cold plasma, when $\Delta p_0 \to 0$, 
i.e. $p_{\pm} \to p$ with $p_{\pm, 0} \to 0$ the  Hamilton function (\ref{eq14-Ham}) reduces to 
\begin{equation}
{\cal H}(E,h)=
\displaystyle{\frac{E^2}{2}+ \gamma_{\rm ph}^2 h  -\beta_{\rm ph} \sqrt{h^2 \gamma_{\rm ph}^4-\gamma_{\rm ph}^2}},
\label{eq15-HT0}
\end{equation}
where $h=\sqrt{1+p^{2}}-\beta_{\rm ph}p$. It can be rewritten in the form of the energy integral,
$E^2/2+\gamma={\rm constant}$. 

\begin{figure}[tbph]
\includegraphics[width=10.4cm,height=4cm]{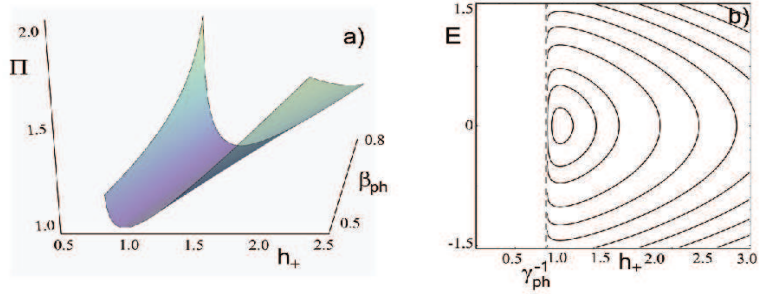}
\caption{ a) Potential function $\Pi(h)$ vs $h$;
 b) Isocontours of the Hamiltonian function in the plane $(E,h)$ for $\beta_{\rm ph}=0.5$. }
\label{fig7}
\end{figure}

The potential  $\Pi(h)=\gamma_{\rm ph}^2 h  -\beta_{\rm ph} \sqrt{h^2 \gamma_{\rm ph}^4-\gamma_{\rm ph}^2}$ 
is plotted in Fig. \ref{fig7}a as a function of $h$ and $\beta_{\rm ph}$.
Isocontours of the Hamiltonian function in the plane $E,h$ for $\beta_{\rm ph}=0.5$ 
are shown in Fig. \ref{fig7}b.

It is easy to see that the electric field equals zero at the maximum of the electron quiver energy. 
This condition yields the result obtained in Ref. \cite{AP} for the maximum value of 
the electric field:  
\begin{equation}
E_{\rm AP}=\sqrt{2 (\gamma_{\rm ph}-1)}.
\label{eq-EmAP}
\end{equation}

For small but finite electron temperature, $\Delta p_0\ll  1/(\beta_{\rm ph}\gamma_{\rm ph})$, 
we obtain (see also Ref. \cite{ESL2})
\begin{equation}
E_{\max}\approx \sqrt{2 (\gamma_{\rm ph}-1)}-\frac{2}{3}\frac{(\beta_{\rm ph}\gamma_{\rm ph})^{3/2}}{\sqrt{(\gamma_{\rm ph}-1)}}\sqrt{\Delta p_0}.
\label{eq-Emx}
\end{equation}

At  $\Delta p_0\to 2\beta_{\rm ph}\gamma_{\rm ph}$ the electric field vanishes because, as mentioned before,
 in this limit the wave 
with arbitrarily small amplitude breaks. 

\begin{figure}[tbph]
\includegraphics[width=14cm,height=4cm]{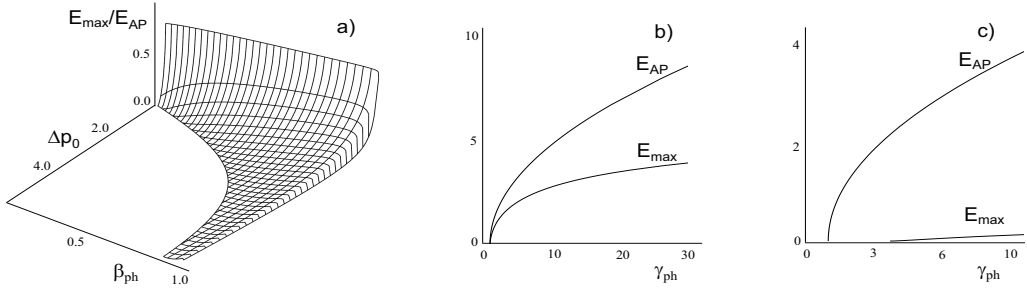}
\caption{Maximum electric field in the breaking wake wave: a)$E_{\max}$  normalized on $E_{\rm AP}$, 
depending on the 
wave phase velocity $\beta_{\rm ph}$ and the width of the electron distribution function, $\Delta p_0$;
The curve in the  $\beta_{\rm ph}$ - $\Delta p_0$ plane where $E_{\max} =0 $   
is given by
 $\Delta p_{0} = 2  \beta_{\rm ph} \gamma_{\rm ph}  $.
b) $E_{\max}$  and $E_{\rm AP}$ v.s. $\gamma_{\rm ph}$ for $\Delta p_0=0.125$.
c) $E_{\max}$  and $E_{\rm AP}$ v.s. $\gamma_{\rm ph}$ for $\Delta p_0=7.5$.}
\label{fig8}
\end{figure}

In Fig. \ref{fig8} we show the maximum electric field, $E_{\max}$, in the breaking wake wave. 
The dependence of this field,
 normalized on $E_{\rm AP}$, on the 
wave phase velocity $\beta_{\rm ph}$ and  on the width of the electron distribution function $\Delta p_0$
is presented in Fig. \ref{fig8} a.  Figures  \ref{fig8} b and c show dependences of  
$E_{\max}$  and $E_{\rm AP}$ on $\gamma_{\rm ph}$ for small and large $\Delta p_0$. 
We see that in the limit $\gamma_{\rm ph} \gg 1$
the difference between $E_{\max}$  and $E_{\rm AP}$ increases according to Eq. (\ref{eq-Emx}).
Figures \ref{fig8} a and c clearly illustrate the above mentioned fact that in the limit
 $\Delta p_0 \to 2 \beta_{\rm ph} \gamma_{\rm ph}$ the value of $E_{\rm max}$ vanishes.

Independently of whether the plasma temperature is finite or vanishes, from Eqs. (\ref{eq14-Ham})
 and (\ref{eq15-HT0}) we obtain that 
the second derivative of the  potential $\Pi$  with respect to $h_+$ ($h$)  
becomes singular at $h_+ = \gamma_{\rm ph}^{-1}$ ($h = \gamma_{\rm ph}^{-1}$), 
which corresponds to the vertical (dashed) singular line in Figs. \ref{fig3}b and \ref{fig6}b.
 In this limit  the Hamiltonian in Eq. (\ref{eq15-HT0}) takes the value
\begin{equation}
\frac{E^2}{2}+\gamma_{\rm ph},   
\quad { \rm while } \quad \gamma (h = \gamma_{\rm ph}^{-1})= \gamma_{\rm ph}. 
\label{eq10-intT0} 
\end{equation}

\begin{figure}[tbph]
\includegraphics[width=6cm,height=10cm]{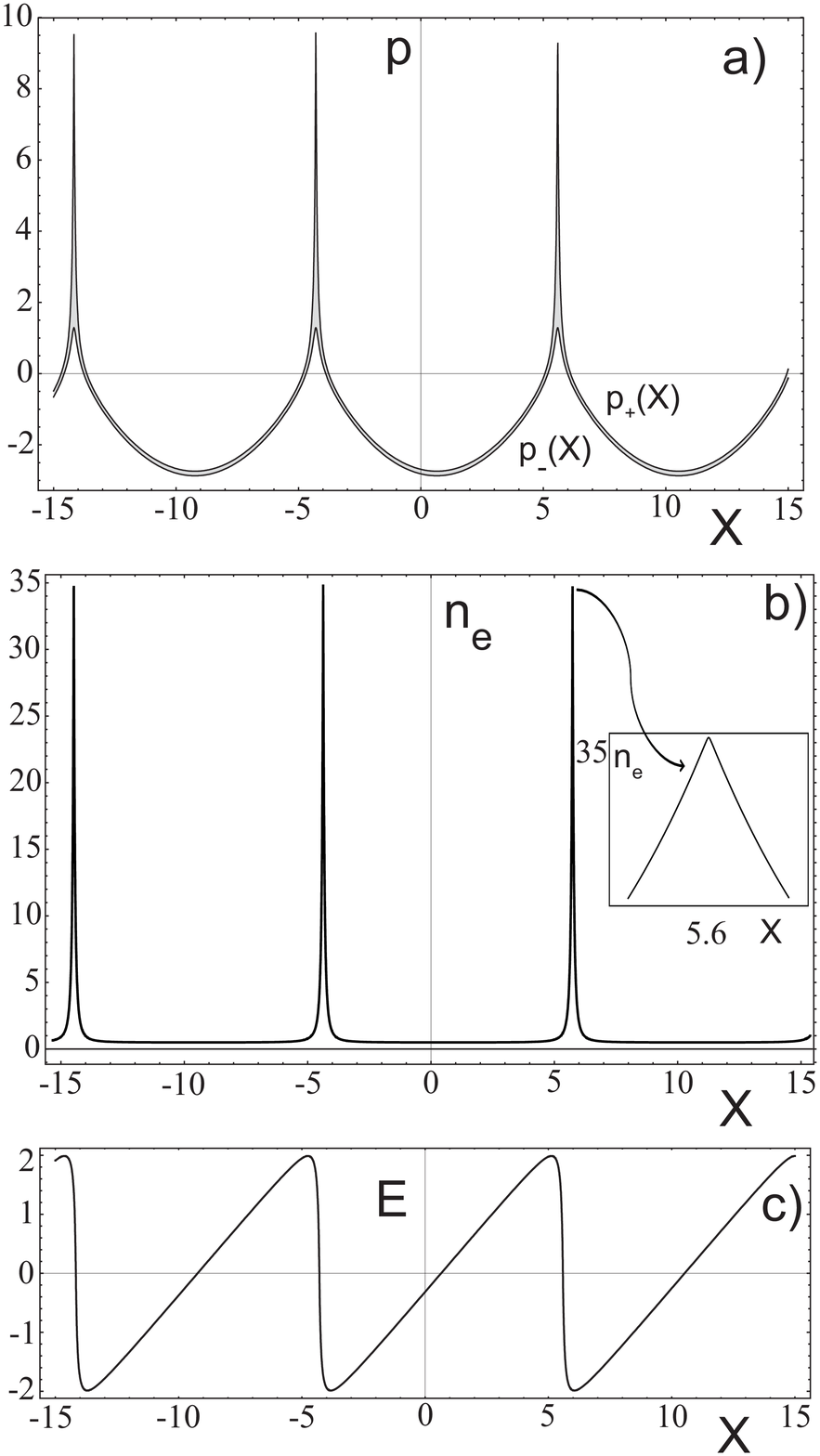}
\caption{Structure of nonlinear wake wave: 
a) electron phase space, 
b) the electron density 
(in the inset the density distribution is shown in the vicinity of the maximum), 
c) electric field
as functions of the coordinate $X$. The normalised wave phase velocity is $\beta_{\rm ph}=0.995$;
the plasma thermal momentum width at $X=15$ is $\Delta p_0=p_{+,0}-p_{-,0}=0.25$;
the maximum of the electric field is $E_m=1.9876$.}
\label{fig9}
\end{figure}

\subsection{Cold wavebreaking limit}

In order to compare the properties of the singularities formed in thermal and cold plasmas we reproduce here
the dependence of the electron momentum and density on the coordinates in the cold 
wavebreaking case (for details see Ref. \cite{PAN}).
In the cold plasma with $p_{+,0} \to 0$ and $p_{-,0} \to 0$, which implies $p_{+}(X)=p_{-}(X)\equiv p$,
equations (\ref{eq8-hpl} -- \ref{eq9-E}) can be reduced to
\begin{equation}
\left(\sqrt{1+p^2}-\beta_{\rm ph} p\right)^{\prime \prime }
=\frac{p}{\beta_{\rm ph}\sqrt{1+p^2}- p}.
\label{eg-m-cold}
\end{equation}%
The solution of this equation can be expressed in terms of elliptic integrals.
In order to analyze
these solutions in the vicinity of the singularity
we note that its right-hand side becomes singular
when the denominator, $\beta_{\rm ph}\sqrt{1+p^2}- p$, tends to zero, i.e., when
the electron velocity $v$ becomes equal to the phase velocity
of the wake wave. In the wake wave, the
singularity is reached at the maximum value of the electron
momentum, $p_{\rm m}=p_{\rm br}$. We assume that the singularity is located 
at the coordinate $X=X_{\rm br}$. 
We consider the wave structure in the vicinity of the singularity, and find that here
the electron momentum depends on $\delta X=X-X_{\rm br}$ as
\begin{equation}
p=p_{\rm br}-\beta_{\rm ph}^{1/3} \gamma_{\rm ph}^2 (3 |\delta X|/2^{1/2})^{2/3}.
\label{eq24-pbr}
\end{equation}%
The electron density tends to infinity as 
\begin{equation}
n\approx n_0 \gamma_{\rm ph}(2^{1/2} \beta_{\rm ph}/3 |\delta X|)^{-2/3}.
\label{eq24-nbre}
\end{equation}

The $2/3$ power behaviour can be recognized in Fig. \ref{fig9}, which presents 
the wakewave generated in a relatively
low temperature plasma with $p_{+,0}\ll p_{\rm br}$. However, from Fig. \ref{fig9}b 
we see that in the very vicinity of the density 
maximum (see inset to Fig. \ref{fig9}b), the dependence of the electron density and momentum on $X$ 
still corresponds to Eq. (\ref{eq23-nebr}),
showing at the wave crest the density profile which can be approximated by the "peakon" 
dependence. The electron distribution width, $p_{\rm{+,br}}-p_{\rm{-,br}}$, near the maximum 
is characterized by the value
\begin{equation}
\Delta p=p_{\rm{+,br}}-p_{\rm{-,br}}\approx \sqrt{2 \beta_{\rm ph} \gamma_{\rm ph}^3 \Delta p_0},
\label{DeltaP-pbr}
\end{equation}%
where we assumed $\gamma_{\rm ph} \Delta p_0\ll 1$.

\section{ Hydrodynamic approach}
{\subsection{Waterbag distribution}}

The system of Eqs. (\ref{eq4-ppl} --\ref{eq6-E}) can be written as a system of hydrodynamic-type equations
\begin{equation}
\partial_t {\mathbb N}+\partial_x{\mathbb J}=0,
\label{App1}
\end{equation}
\begin{equation}
\partial_t {\mathbb P}+\partial_x{\mathbb G}=-E,
\label{App2}
\end{equation}
\begin{equation}
\partial_t E+{\mathbb V}\partial_x E={\mathbb V}
\label{App3}
\end{equation}
for the electron density 
\begin{equation}
{\mathbb N}(x,t)=\int{f_e(p,x,t) dp}=\frac{1}{\Delta p_0}(p_{+}-p_{-}),
\label{App4}
\end{equation}
average momentum
\begin{equation}
{\mathbb P}(x,t)=\frac{1}{{\mathbb N}}\int{p f_e(p,x,t) dp}
=\frac{p_{+}^2-p_{-}^2}{2(p_+-p_-)}=\frac{1}{2}(p_{+} + p_{-}),
\label{App5}
\end{equation}
and electric field $E$.
Here 
\begin{equation}
{\mathbb J}(x,t)=\int{\frac{p}{\sqrt{1+p^2}} f_e(p,x,t) dp}=\frac{1}{\Delta p_0}(\gamma_{+}-\gamma_{-}),
\label{App6}
\end{equation}
with $\gamma_{\pm}={\sqrt{1+p_{\pm}^2}}$,
\begin{equation}
{\mathbb V}(x,t)=\frac{1}{\mathbb N}\int{\frac{p}{\sqrt{1+p^2}} f_e(p,x,t) dp}
=\frac{p_{+} + p_{-}}{\gamma_{+}+\gamma_{-}},
\label{App7}
\end{equation}
and
\begin{equation}
{\mathbb G}(x,t)=\frac{\displaystyle{\int{p f_e(p,x,t) dp}}}{\displaystyle{\int{\frac{p}{\sqrt{1+p^2}} f_e(p,x,t) dp}}}=\frac{1}{2}(\gamma_{+} + \gamma_{-}),
\label{App8}
\end{equation}
These functions are related to each other as
\begin{equation}
{\mathbb V}= \frac{{\mathbb P}}{\mathbb G}=\frac{{\mathbb J}}{\mathbb N}
\label{App9}
\end{equation}
and
\begin{equation}
{\mathbb G}= 
\sqrt{\frac{1}{4}+\left(\frac{{\mathbb P}}{2}+\frac{{\mathbb N}\Delta p_0}{4}\right)^2}
+\sqrt{\frac{1}{4}+\left(\frac{{\mathbb P}}{2}-\frac{{\mathbb N}\Delta p_0}{4}\right)^2}.
\label{App10}
\end{equation}

In the case of a wave travelling with constant velocity $c \beta_{\rm ph}$, 
the functions ${\mathbb N}, {\mathbb P}, {\mathbb G}$ and $E$ depend 
on the variable $X=x-\beta_{\rm ph} t$  and  Eqs. (\ref{App1} -- \ref{App3}) can be reduced to
\begin{equation}
\left({\mathbb G}-\beta_{\rm ph}{\mathbb P}\right)^{\prime \prime }= 
-\frac{{\mathbb P}}{\beta_{\rm ph}{\mathbb G}-{\mathbb P}},
\label{App11}
\end{equation}
\begin{equation}
{\mathbb N}= \frac{\beta_{\rm ph}{\mathbb G}}{\beta_{\rm ph}{\mathbb G}-{\mathbb P}},
\label{App12}
\end{equation}
where a prime denotes differentiation with respect to $X$.
Equation (\ref{App11}) 
looks  identical to Eq. (\ref{eg-m-cold}) which describes the wave break
at $p/\gamma \to \beta_{\rm ph}$ and the formation of a  singularity  in the electron density, $n\to\infty$, 
 according to
Eq. (\ref{eq24-nbre}). However, due to the nonlinear dependence of ${\mathbb G}$ on ${\mathbb P}$ given
by relationships (\ref{App10}) and (\ref{App12}) the character of the singularity   changes and  becomes 
of the type
described in Sec. \ref{cres}. In particular, we can see that the condition for
 the denominator in the r.h.s. 
of Eq. (\ref{App11}) to vanish implies that ${\mathbb V}=\beta_{\rm ph}$. This condition can be rewritten as 
\begin{equation}
\beta_{\rm ph}=\frac{p_{+} + p_{-}}{\gamma_{+}+\gamma_{-}}=
\frac{p_{+}}{\gamma_{+}} \left(\frac{1+ p_{-}/p_{+}}{1+\gamma_{-}/\gamma_{+}}\right).
\label{App13}
\end{equation}
Assuming that in this limit $p_{-}=p_{+}+\delta p$ with $\delta p/p_{+}\ll 1$, we can easily find that the
condition of "hydrodynamic type wave break"(\ref{App13}) used in \cite{KM} is equivalent to 
\begin{equation}
\beta_{\rm ph}=\frac{p_{+} + p_{-}}{\gamma_{+}+\gamma_{-}}=
\frac{p_{+}}{\gamma_{+}} \left(1-\frac{\delta p}{p_{+}^3}\right),
\label{App14}
\end{equation}
which requires 
${p_{+}}/{\gamma_{+}}>\beta_{\rm ph}$, i.e. the waterbag description in the adopted  limit 
of a  stationary nonlinear wave propagating
with constant velocity is no longer valid.

{\subsection{Nonrelativistic limit}}

In the nonrelativistic limit Eqs. (\ref{App1}) and (\ref{App2}) take the form (see Ref. \cite{RCD})
\begin{equation}
\partial_t{\mathbb N}+\partial_x\left({\mathbb N}{\mathbb V}\right)=0,
\label{App15}
\end{equation}
\begin{equation}
\partial_t{\mathbb V}+{\mathbb V}\partial_x{\mathbb V}=-E-\frac{\Delta p_0^2}{8}\partial_x {\mathbb N}^2,
\label{App16}
\end{equation}
which corresponds to a gasdynamics system where  the pressure depends  on the gas density as 
\begin{equation}
P=\frac{P_0}{{\mathbb N}_0^3} {\mathbb N}^3
\label{App17}
\end{equation}
with $P_0$=const.

For a wave travelling with constant velocity, $ \beta_{\rm ph}$,  
 we obtain 
\begin{equation}
\left[\frac{1}{2}\left({\mathbb V}-\beta_{\rm ph}\right)^2-\frac{\beta_{\rm ph}^2\Delta p_0^2}
{8\left({\mathbb V}-\beta_{\rm ph}\right)^2}\right]^{\prime \prime }= 
-\frac{{\mathbb V}}{\beta_{\rm ph}-{\mathbb V}}.
\label{App18}
\end{equation}
The singular points of this equation correspond to
\begin{equation}
{\mathbb V}_1=\beta_{\rm ph} \quad {\rm and} 
\quad {\mathbb V}_{2,3}=\beta_{\rm ph}\pm \sqrt{\frac{\beta_{\rm ph}\Delta p_0}{2}} .
\label{App19}
\end{equation}

 We see that  the points ${\mathbb V}_1$ and ${\mathbb V}_2$ lay beyond the applicability 
range of  the waterbag model,  while a  wave with ${\mathbb V}_{\max} \to {\mathbb V}_3$ 
is qualitatively described by Fig. \ref{fig9} with
maximum  density $n_{\max}=\sqrt{2 \beta_{\rm ph}/ \Delta p_0}$ and electric field 
$E_{\max}\approx \beta_{\rm ph}- \sqrt{\beta_{\rm ph} \Delta p_0/2}$.

{\subsection{Ultrarelativistic limit}}

This case corresponds to the limit ${\mathbb P} \gg 1$. Expanding Eq.(\ref{App10}) into
series of  the  small parameter ${\mathbb N}\Delta p_0/{\mathbb P}\ll 1$ we obtain
\begin{equation}
{\mathbb G}\approx \sqrt{1+{\mathbb P}^2}+\frac{{\mathbb N}^2\Delta p_0^2}{8(1+{\mathbb P}^2)^{3/2}}.
\label{App20}
\end{equation}
Using Eq. (\ref{App12}) for the electron density ${\mathbb N}$ 
we find from Eqs. (\ref{App11}) and (\ref{App20})
\begin{equation}
\left[\sqrt{1+{\mathbb P}^2}-\beta_{\rm ph}{\mathbb P}
+\frac{\beta_{\rm ph}^2\Delta p_0^2}
{8\sqrt{1+{\mathbb P}^2}(\beta_{\rm ph}\sqrt{1+{\mathbb P}^2}-{\mathbb P})^2}\right]^{\prime \prime }= 
-\frac{{\mathbb P}}{\beta_{\rm ph}\sqrt{1+{\mathbb P}^2}-{\mathbb P}}.
\label{App21}
\end{equation}

The singular points of this equation, written in terms of the average velocity 
${\mathbb V}={\mathbb P}/\sqrt{1+{\mathbb P}^2}$,
are given by  
\begin{equation}
{\mathbb V}_1\approx \beta_{\rm ph} \quad {\rm and} \quad 
{\mathbb V}_{2,3}\approx \beta_{\rm ph} \pm \sqrt{\frac{\beta_{\rm ph}\Delta p_0}{2\gamma_{\rm ph}^3}}
\label{App22}
\end{equation}
with ${\mathbb V}_{3}$ corresponding to the wake wave breaking.
This yields for maximum  density $n_{\max}\approx\sqrt{2 \beta_{\rm ph} \gamma_{\rm ph}^3/ \Delta p_0}$ 
and for the  electric field 
$E_{\max}\approx \sqrt{2(\gamma_{\rm ph}-1)}- (2/3)\gamma_{\rm ph}\sqrt{\beta_{\rm ph}\Delta p_0}$   
in agreement with Eqs. (\ref{eq17-nebrrl}) and (\ref{eq-Emx}).

\bigskip

\section{Computer simulation of the plasma wave breaking in thermal plasmas}

During the irradiation of underdense plasma targets by high-power laser 
pulses, the light within  the pulse 
generates a finite amplitude wake wave whose parameters depend, in particular, on the plasma
temperature and on the interaction geometry. A thorough study of 
these effects require computer simulations.
We performed parametric studies of the laser pulse interaction with underdense targets 
using 
a two-dimensional (2D3P) particle-in-cell (PIC) code \cite{NPIC}.


Here the 
 effects  of the finite electron temperature have been taken into account in three limiting cases.
In the  first case  the initial electron temperature has been assumed to be equal to zero.
In the second case thermal effects have been modelled by the electron distribution corresponding 
to the  initial waterbag distribution function with a  temperature equal to $100 $eV. 
In the third case, the initial electron distribution was Maxwellian with the same temperature. 
In both the cases of the waterbag and Maxwellian distributions, 
the total electron energy is the same, 
i.e., average energy for the waterbag case,
 \begin{equation}
 \left<m_ec^2(\gamma-1)\right> \approx \left<\frac{p^2}{2\,m_e}\right>
 =\displaystyle{
 \frac{\int_{0}^{\Delta p_0}(p^2/2\,m_e) p^2dp}{\int_{0}^{\Delta p_0} p^2dp}
 =\frac{3 \Delta p_0^2}{40 m_e}},
\end{equation} 
 is set to be equal to that in the case of a Maxwellian distribution, $<p^2/2\,m_e>=(3/2)k_BT$. 
Here we assumed that $p/mc \ll 1$.
  
In these simulations, the laser pulse has a normalized amplitude of $a_0=eE_0/m_e\omega c=4.6$,
 a wavelength of $\lambda=2 \pi c/\omega=0.8\,\mu$m, focused onto a  spot of the size of $13\mu$m, and  duration of 16 fs. 
The plasma density equals $4\times 10^{19}$cm$^{-3}$. 
The width of the simulation box is equal to  $50\times 65 \lambda^2$.
The mesh size is $\Delta x=\lambda/160$ with 30 particles per cell.

\begin{figure}[tbph]
\includegraphics[width=12cm,height=9cm]{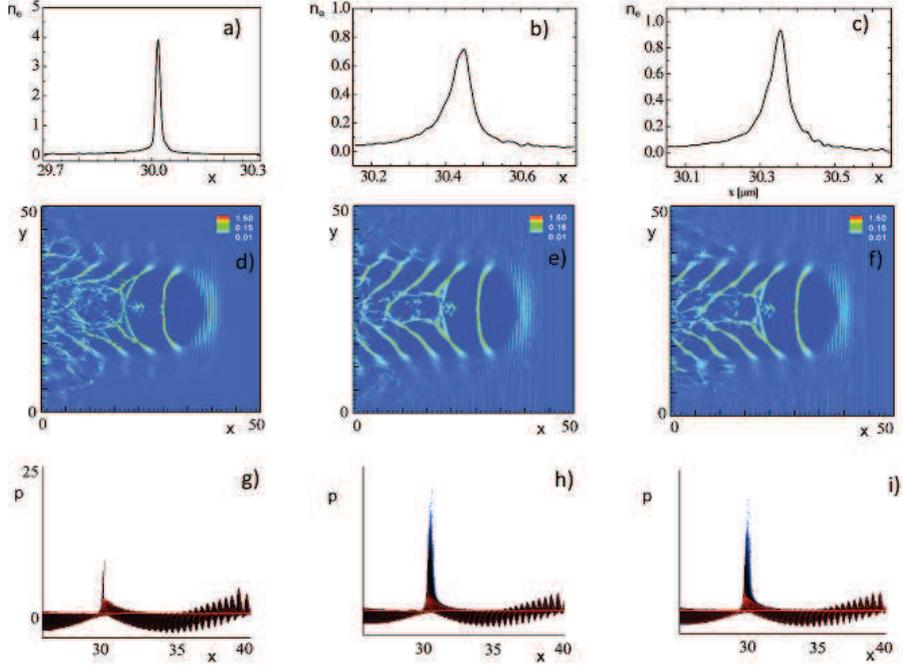}
\caption{Results of 2D-PIC simulations of the nonlinear wake wave generation in thermal plasmas: 
electron density and phase space in a cold plasma (a,d,g); 
in the plasma described by the waterbag distribution with a temperature of 100 eV (b,e,h);
in the plasma described by the Maxwellian distribution with a temperature of 100 eV (c,f,i).}
\label{fig10}
\end{figure}

Simulation results for the parameters of interest are shown in Fig. \ref{fig10}. 
Here the $x$ coordinate
is measured in $1\,\mu$m,  
the electron momentum $p$ is normalized on $m_e c$, and the 
density is normalized on the critical density $n_{cr}=m_e \omega^2/4 \pi e^2$.
The figures are plotted for the time when the highest density is reached, 
which is  350 fs, 310fs, 310fs for zero-temperature, waterbag, and Maxwellian distribution, respectively.  
In the cold plasma case, the electron density distribution in the first maximum of the 
breaking wake wave takes a  cusp-like form (see Fig. \ref{fig10} a.
In Figs. \ref{fig10} b and c we see that the finite temperature effects lead to a decrease
of the maximum  electron density in the 
breaking wake wave, to the  broadening of the maximum  and to the formation of  
peakon-like structures for both
the waterbag and the  Maxwellian distributions. Note here more efficient electron injection
in the finite temperature plasma compared with the cold plasma case.

\section{Above the wavebreaking limit}

\subsection{Maximal electric field}

The limiting electric field given by Eq. (\ref{eq-EmAP}) 
corresponds  to a  stationary Langmuir wave for which the electron 
quiver energy is below $m_ec^2\gamma_{\rm ph}$.
When the Langmuir wave is excited by a short laser pulse its amplitude 
and its  phase velocity depend on the
plasma density and on the laser pulse intensity \cite{ESL1}. Propagating in an underdense plasma, 
an intense laser pulse can accelerate plasma electrons
longitudinally up to the energy $m_ec^2(1+a^2/2)$. 
In a cold plasma the wavebreaking condition 
corresponds to $(1+a^2/2)=\gamma_{\rm ph}$, where $\gamma_{\rm ph}=(n_{cr}/n_0)(1+a^2)^{1/4}$ 
is the Lorentz 
gamma-factor calculated for the wake wave phase velocity  which is equal to the laser pulse group velocity.
Here the  dependence of the 
electromagnetic wave group velocity  on its amplitude is taken into account  according to \cite{AP}. 
This yields the wake wave 
breaking threshold in terms of the driver laser pulse amplitude \cite{EK}
\begin{equation}
a>\left(4 \frac{n_{cr}}{n_0}\right)^{1/3}.
\label{eqEK}
\end{equation}

In general, the laser pulse amplitude in a plasma is different from its value in   vacuum 
due to the laser pulse self-focusing and self-channelling \cite{RSF}. The laser pulse amplitude 
inside the self-focusing channel relates to the laser power ${\cal P}$ and plasma density as \cite{SSB}
\begin{equation}
a_0^3>8 \pi \frac{{\cal P}}{{\cal P}_c}\frac{n_0}{n_{cr}},
\label{eqSSB}
\end{equation}
where ${\cal P}_c=2m_e^2c^5/e^2\approx 17$GW.

Using Eqs. (\ref{eqEK}) and (\ref{eqSSB}) we obtain  the wake wave breaking threshold:
\begin{equation}
{\cal P}>\frac{{\cal P}_c}{2 \pi}\left(\frac{n_{cr}}{n_0}\right)^2.
\label{eqWWBTH}
\end{equation}

For example, from Eqs. (\ref{eqEK}) and (\ref{eqWWBTH}) we find that if a  laser pulse 
of the wavelength $\lambda=0.8\mu$m, 
for which $n_{cr}\approx 2\times 10^{21}{\rm cm}^{-3}$, propagates in a plasma 
 with   density  $n_0=2\times 10^{19}{\rm cm}^{-3}$, the wavebreaking threshold is reached for 
${\cal P}>30\,$TW and 
$a_0=7.3$, i.e. for   a  laser intensity of the order of $7\times 10^{19}\,{\rm W/cm}^{2}$.

A  laser pulse with power  larger than  that given by the r.h.s of Eq. (\ref{eqWWBTH}), 
causes the wake wave to break  in the first period, with the electric field well above 
the limiting value given 
by Eq. (\ref{eq-EmAP}) and with  a number of electrons piled up in the singularity region 
much larger than in the stationary case described by Eqs. (\ref{eq15-nebr}) and (\ref{eq24-nbre}). 
This fact has important consequences for determining the laser wakefield acceleration 
scaling \cite{LWFA,ESL1}.
 
A wakewave with  an amplitude above the wave break threshold is transient and forms
  a region with   multi-stream electron motion. The multi-stream motion region expands in the forward 
direction at a relative velocity $dX/dt\approx c(1-\beta_{\rm ph})\approx c/2\gamma^2_{ph}$. 
Since in the 
limit $\gamma_{\rm ph} \gg 1$ this velocity is low, the region with a  large electric field 
(and with a large number of  electrons) can exist for a  substantially long time, 
which is of the order of the charged particle acceleration time, 
$ t_{\rm acc}=2 \lambda_w \gamma^2_{ph}/c$. Here $\lambda_w$ is the wake wave wavelength.

The structure of the wake wave both below and above the wavebreaking limit can be revealed from
the phase plane pattern presented in Figs. \ref{fig4} b and \ref{fig7} b.
The stationary (periodic) waves correspond to the bound trajectories in the phase plane shown
in Figs. \ref{fig4} b and \ref{fig7} b. The stationary breaking wave is described by a last closed trajectory 
touching the vertical (dashed) singular line, corresponding to $h_{+}=\gamma_{\rm ph}^{-1}$, 
in this figure.
 
In the  vicinity of  the singular line in Fig. \ref{fig4}b  
 the Hamiltonian function (\ref{eq14-Ham}) with the potential in the form given by Eq. (\ref{eq14-POTsym})
can be expanded in series of $\delta h_{+}=h_{+}-\gamma_{\rm ph}^{-1}$ as 
\begin{equation}
{\cal H}(E,h)=
\displaystyle{\frac{E^2}{2}+
\left. \Pi\right\vert_{h=\gamma_{\rm ph}^{-1}}+\left. \frac{d \Pi}{d h}\right\vert_{h=\gamma_{\rm ph}^{-1}}
\delta h}+ ...\, .
\label{POTsymexp-thrm}
\end{equation}
Here we assume a  symmetric electron distribution at $X=X_0$, where $p_{+,0}= -p_{-,0}$.

For a finite temperature plasma  in the vicinity of the singularity we find
\begin{equation}
{\cal H}(E,h)=
\frac{E^2}{2}- 
\gamma_{\rm ph}^{2}
\left[
\left(
\beta_{\rm ph}^2 +
\frac{2\beta_{\rm ph}}{\gamma_{\rm ph} \Delta p_0}
\right)^{1/2}-1 
\right]
 \delta h_+,
\label{eq15-HT0hexpt}
\end{equation}
where the constant  term   
$\gamma_{\rm ph} - W(1 +\beta_{\rm ph} \gamma_{\rm ph} \Delta p_0)/(2 \Delta p_0)$ has been dropped. 
At the wavebreaking threshold, the value of the Hamiltonian 
 ${\cal H}=H_1 =0 $  and the electric field $E$ tends to zero at $\delta h_+ \to +0$ as
\begin{equation}
E=\gamma_{\rm ph} \left[\left(\beta_{\rm ph}^2 +\frac{2\beta_{\rm ph}}{\gamma_{\rm ph} \Delta p_0}\right)^{1/2}-1 
\right]^{1/2} \sqrt{2\delta h_+},
\label{eq16-Eexpt}
\end{equation}
For $H_1>0$ the electric field $E_1$  at  wave break,  $h_+=\gamma_{\rm ph}^{-1}$,  
is given by  $E_1=\sqrt{ 2H_1}$. 

The quantity of the Hamiltonian $H_1$ is determined by the parameters 
of the laser pulse driver generating the wake wave. In the limit of large laser amplitude, $a\gg 1$, assuming that the laser pulse has 
an optimal 
duration, $\tau_{\rm las}\approx a/2\omega_{\rm pe}$, 
we can find that $H_1\approx 4\pi n_0 m_e c^2 a^2$, i.e.
 the maximal electric field is given by $E_{\rm max}\approx a \sqrt{8 \pi n_0 m_e c^2}$.

For $H_1<0$, 
the wave breaking condition is not reached, 
 and  the electric field vanishes at 
 \begin{equation}
\delta h_+ = \delta h_{+1}=\frac{H_1}{
 \gamma_{\rm ph}^{2}
 [\beta_{\rm ph}^2 +(2 \beta_{\rm ph})/(\gamma_{\rm ph} \Delta p_0)]^{1/2}-\gamma_{\rm ph}^{2}] }
\end{equation}
 as
\begin{equation}
E =\gamma_{\rm ph} \left[\left(\beta_{\rm ph}^2 +\frac{ 2 \beta_{\rm ph}}{\gamma_{\rm ph} \Delta p_0}\right)^{1/2}-1 \right]^{1/2}  (2{\delta h}-2{\delta h_{+1}})^{1/2}.
\label{eq16-Eexp-1t}
\end{equation}

In the cold plasma limit the Hamiltonian (\ref{eq15-HT0})  expansion in the vicinity of the singularity 
 has a different behaviour:
\begin{equation}
{\cal H}(E,h)=
\displaystyle{\frac{E^2}{2}-\beta_{\rm ph}\gamma_{\rm ph}^{3/2}\sqrt{2\delta h}},
\label{eq15-HT0exp}
\end{equation}
where the constant $\gamma_{\rm ph} $  term has been dropped. 
At the wavebreaking threshold, the value of the Hamiltonian 
 ${\cal H}=H_1 =0 $  and the electric field $E$ tends to zero at $\delta h \to +0$ as
\begin{equation}
E=[\beta_{\rm ph}(2 \gamma_{\rm ph})^{3/2}]^{1/2}(\delta h)^{1/4},
\label{eq16-Eexp0}
\end{equation}
For $H_1>0$ the electric field $E_1$  at  wave break,  $h=\gamma_{\rm ph}^{-1}$,  
is given by  $E_1=\sqrt{ 2H_1}$. For $H_1<0$, 
the wave breaking condition is not reached, 
 and  the electric field vanishes at $\delta h = \delta h_1=H_1^2/(2\beta_{\rm ph}^2 \gamma^3_{ph})$ as
\begin{equation}
E =[\beta_{\rm ph}(2\gamma_{\rm ph})^{3/2}]^{1/2}(\sqrt{\delta h}-\sqrt{\delta h_1})^{1/2}.
\label{eq16-Eexp-1}
\end{equation}

In the limit of a relatively low plasma temperature $\Delta p_0 \ll \beta_{\rm ph} \gamma_{\rm ph}$
in order to estimate the maximum electric field we can use the Hamiltonian in the form given by 
Eq. (\ref{eq15-HT0}). 
In this limit the maximum electric field, 
$E_{\max}=\sqrt{2\left(\Pi(h_{\rm m}) - \Pi(h_{\rm E})\right)}$ with $h_{\rm E}=1$ and $h_{\rm m}$
determined by the maximal electron quiver energy in the wake (see Fig. \ref{fig6}),  is given by
\begin{equation}
E_{\max}=\sqrt{2\left(\gamma_{\rm m} - 1\right)}.
\label{eq-Eabwbrmax}
\end{equation}
The electric field at  the wake wave breaking point is equal to 
$\sqrt{2\left(\Pi(h_{\rm m})-\Pi(\gamma_{\rm ph}^{-1}) \right)}$, which 
in the limit of a relatively low plasma temperature yields
\begin{equation}
E_{\rm br}=\sqrt{2\left(\gamma_{\rm m} - \gamma_{\rm ph}\right)}.
\label{eq-Eabwbr}
\end{equation}
For a  wake wave with a  large enough amplitude, when $\gamma_{\rm m} \gg \gamma_{\rm ph}$, both the maximum
electric field and the electric field at the breaking point can be substantially larger than the
electric field in the stationary wake wave given by Eq. (\ref{eq-Emx}). 

As we see in Figs. \ref{fig4} and \ref{fig7} 
 in the  regime under the consideration the injected electrons appear in the region 
 $h>h_{\rm br}=\gamma_{\rm ph}^{-1}$ with  a large accelerating 
electric field.

  At wave break electrons  are injected   into the region $h>h_{\rm br}=\gamma_{\rm ph}^{-1}$ 
  where there is a large accelerating 
electric field,  as seen  in Figs. \ref{fig4} and \ref{fig7}.
Note that since the electric field at the breaking point does not vanish, the type of the singularity
that is formed in the electron momentum and density distributions changes. 
In a finite temperature plasma the electron 
density in the vicinity of the singular point is determined by Eqs. (\ref{eq19-npsing}) 
and (\ref{eq19-psing}).
From Eq. (\ref{eq19-psing}) we find
\begin{equation}
\delta p_{+}=-\sqrt{\gamma_{\rm m}^3 n_{e,\rm br} \delta X^2+ 2 \gamma_{\rm m}^3 E_{\rm br} \delta X}
\label{eq-pplabove}
\end{equation}
where 
$E_{\rm br}$ is given by Eq. (\ref{eq-Eabwbr}) and  it is assumed that $\delta X>0$.
Inserting Eq. (\ref{eq-pplabove})  into  Eq. (\ref{eq19-npsing}) 
we find that for $E_{\rm br}\neq 0$   
the electron density near the singularity behaves for $ \delta X \to +0$ as  
\begin{equation}
n_e=n_{e,\rm br}-\frac{1}{\Delta p_0} \sqrt{2 \gamma_{\rm m}^3 E_{\rm br} \delta X}.
\label{eq-neabove}
\end{equation}

In the limit of cold plasma, $\Delta p_0 \to 0$, Eq. (\ref{eg-m-cold}) yields (see also \cite{PAN})
\begin{equation}
(\delta p^2)^{\prime \prime}=-\frac{2 \beta_{\rm ph} \gamma_{\rm ph}^6}{\delta p}.
\label{eq-pabovecold}
\end{equation}
Multiplying the left- and right-hand sides of this equation by  $(\delta p^2)^{\prime}$ and 
integrating over  $X$, we obtain
\begin{equation}
\delta p \delta p^{\prime}=\sqrt{2 \gamma_{\rm ph}^3 E_{\rm br}-2 \beta_{\rm ph} \gamma_{\rm ph}^6 \delta p}.
\label{eq-pabovecoldint}
\end{equation}
For $E_{\rm br}\neq 0$ the main term in the expansion of  the solution of 
Eq. (\ref{eq-pabovecoldint}) for  $\delta X \to +0$ is
\begin{equation}
\delta p =-(8 \gamma_{\rm ph}^3 E_{\rm br})^{1/4}\, \sqrt{\delta X}.
\label{eq-sqrtcoldsing}
\end{equation}
Using this relationship we find that   in the vicinity of the 
singularity  the density  depends on $\delta X$ as
\begin{equation}
n_e \approx \frac{ \beta_{\rm ph} \gamma_{\rm ph}^{9/4}}
{(8 \gamma_{\rm ph}^3 E_{\rm br})^{1/4}\, \sqrt{\delta X}}.
\label{eq-neabovecold}
\end{equation}
If instead $E_{\rm br}=0$,   the electron momentum and density  are given by 
Eqs. (\ref{eq21-psinX}, \ref{eq23-nebr}) 
for 
$\Delta p_0 \neq 0$ and by  Eqs. (\ref{eq24-pbr}, \ref{eq24-nbre})  for $\Delta p_0 = 0$, 
respectively.

\subsection{Results of simulations with the 1-D Vlasov code}


The Vlasov-Poisson system is solved for the electron 
distribution function, 
$f_{e}(x,v,t)$, with the numerical scheme described in Ref. \cite{FRCA}, 
limiting our study to the 1D-1V case.
The equations are normalized by using the following characteristic quantities: 
the charge $e$ and the electron mass $m_{e}$. The electron density is
normalized on the density of ions $n_{0}$, which are assumed to be at rest. 
Time and space coordinate are normalized on the inverse Langmuir frequency $\omega_{pe}^{-1}$
and on  
the Debye length $\lambda_{D} = \sqrt{T_{e} / 4 \pi n_{e} e^2}$, respectively.
The electron velocity is normalized on the electron thermal velocity 
$v_{th,e} = \lambda_{D} \omega_{pe} = \sqrt{T_{e} / m_{e}}$ 
and the electric field is measured in units $ m_{e} v_{th,e} \omega_{pe} / e$. 
Then, the dimensionless equations 
read:
\begin{equation} 
\label{eq:vlasovelectrons}
\partial_t f_{e} + v \partial_x f_{e} - (E+E_{\rm ext}) \partial_v f_{e}= 0 
\end{equation}
for the electron distribution function and
\begin{equation} \label{eq:poisson}
\partial_{xx} \phi = \int f_{e} dv - 1
\end{equation}
for the electrostatic potential, $\phi$ with $E=-\partial_x \phi$.
Here $E_{\rm ext}$ is an external driver added to the Vlasov equation 
that can be switched on or off during the run. 
The electron distribution function is discretized in 
space for $0 \leq x < L_{x}$, with $L_{x} = 500\ \lambda_{D}$ 
the total box length, with a resolution of $dx = 0.1 \lambda_{D}$.
The electron velocity grid ranges over $-80 \ v_{\rm th,e} \leq v \leq +580  v_{\rm th,e}$, 
with a resolution of $\Delta v = 0.0533\ v_{\rm th,e}$. 
Finally, periodic boundary conditions are used in the spatial direction.

The plasma is  initially homogeneous with waterbag electron distribution, which is modelled by 
the super-Gaussian function
$f_{e}(x,v) = \exp({-v^8})/{[2 \Gamma(9/8)]} $ with $\Gamma (x)$ the Euler gamma function \cite{AS}.

Added to the Vlasov equation Eq.~\ref{eq:vlasovelectrons}) external driver~$E_{\rm ext}$ is given by
$E_{\rm ext}  (x,t) = 0$  if   $t < t_1$ or $t > t_2$, 
$E_{\rm ext}  (x,t) =  - 2A (x_{\rm g}/L) \exp{(- x_{\rm g}^2 )} \left[1 - \exp{(- 2 (t - t_1) )} \right]$
for $t_1 \leq t \leq t_2$. Here $x_{\rm g} = (x - x_0 - v_{\rm ph} t ) / L$ 
with $L = 0.0625$, $v_{\rm ph} = 10$, $A = 150$, $t_1 = 1$ and $t_2 = 3$.

\begin{figure}[tbph]
\includegraphics[width=10cm,height=8cm]{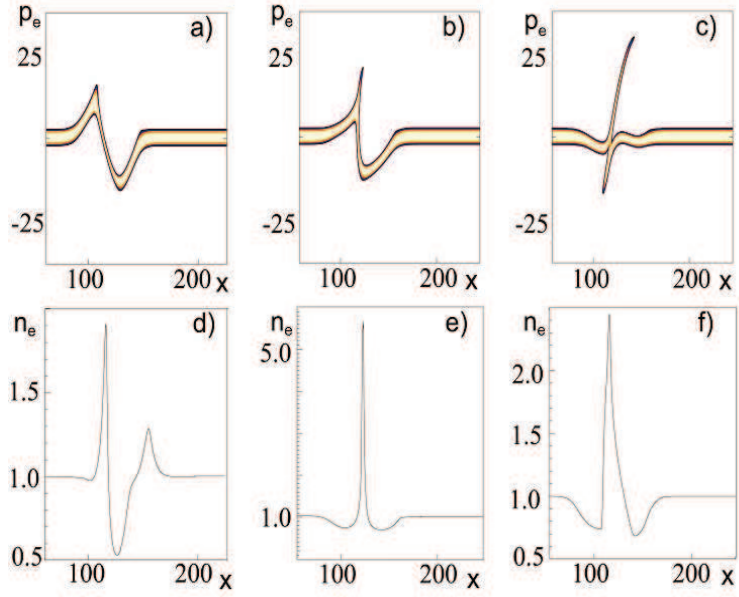}
\caption{Results of 1D-Vlasov simulations of nonlinear wake wave breaking in thermal plasmas: 
electron phase plane in the plasma described by the waterbag distribution (a,b,c)
and
electron density profile (d,e,f) for $t=5,6,7$, respectively. }
\label{fig11}
\end{figure}

The results of the Vlasov simulations of nonlinear wake wave breaking in thermal plasmas 
are presented in Fig.\ref{fig11}, where we show the electron phase space 
  and electron density profile for $t=5,6,7$.
  The electron momentum is normalized 
on $m_e v_{\rm th,e}$ and density on the ion density $n_0$.
  As we see in Fig. \ref{fig11} a, at time $t=5$, 
  when the electron velocity reaches $v_{\rm ph}$, the wake wave
 starts to break with the singularity corresponding 
 to above discussed the "$\Lambda$-type breaking", which 
 results in the narrow density spike shown in in Fig. \ref{fig11} d. 
 The electron multistream region is formed at $t>5$ as seen in Fig. \ref{fig11} b. 
 Due to the momentum conservation the wake wave experiences
 a recoil leading to a  slowing down of its propagation velocity and 
 to a backward acceleration of the electrons in
the region localized ahead of the wavebreaking point 
and to piling up the electron density, which make the electron
density spike to be more narrow with high electron density inside (Fig. \ref{fig11} b, e). 
At $t=6$ and $7$ the electron phase space evolves
into the the structure, which can be called "the $N$-type breaking" (Fig. \ref{fig11} c, f). 
 Later  the multistream motion region becomes wide and the electron density
maximum becomes broader. 

\subsection{Simple model}

\begin{figure}[tbph]
\includegraphics[width=10cm,height=4cm]{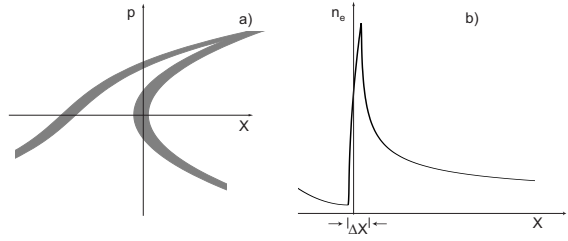}
\caption{Simple model of the wave breaking:
a) electron phase plane, b) electron density. }
\label{fig12}
\end{figure}

Consideration of Fig. \ref{fig11} showing
 the singularity structures formed during and after the wave breaking leads to 
the formulation of a simple model within whose framework we can explain analytically the 
main features seen in the electron density distribution.
As we may see from Fig. \ref{fig11} b the "$\Lambda$-type breaking" in the phase plane, $p,X$,
can be locally approximated by a superposition of two finite width stripes of parabolic and cubic form
as is illustrated in Fig. \ref{fig12} a.
In other words, the waterbag distribution function is constant within the regions marked by the curves 
$p_{\pm}(X)$ given by equations
\begin{equation} 
\label{eq:w-b-parab}
p_{\pm}^2/2= X\pm \Delta X/2
\end{equation}
in the part corresponding to the parabolic behaviour
and 
\begin{equation} 
p_{\pm}^3-r \, p_{\pm}= X-X_c\pm \Delta X/2
\label{eq-cubics}
\end{equation}
for the cubic part.
The parameters $r$ and $X_c$ provide the overlapping of these two stripes at large $p$, with 
$\Delta X$ being the distribution width at $p=0$. 

In order to parametrize these dependences we consider the electron motion in the frame of reference,
where the singularity region is at rest. The parabolic stripe here can be described using an approximation 
of the integral of motion, $m_e c \gamma =m_e c \gamma_0+e E X $, in the vicinity of the reflection point,
where $p \to 0$, i.e. $p^2/2 =m_e c (\gamma_0-1)+e E X $. We find that $\Delta X$ 
in Eq. (\ref{eq:w-b-parab}) is proportional 
to the width $\Delta p_0$ of the initial momentum  distribution and inversely
 proportional to the electric 
field reflecting back the electrons in the wave breaking region: 
$\Delta X=(m_e c/eE)(\gamma_{+,0}-\gamma_{-,0}) \approx \Delta p_0/eE$. In the laboratory frame of reference 
the distribution width is approximately $2 \gamma_{\rm ph}$ times narrower.

The electron density can be calculated as the  area within the $p_{\pm}(X)$ curves. Its part corresponding
to the parabolic curves is equal to
\begin{equation} 
\label{eq:w-b-n_e-par}
n_e(X)=n_0
\frac{2\sqrt{2 e E m_e c}}{\Delta p_0}
\left[ 
\theta \left(
X+\frac{\Delta X}{2}
\right)
\sqrt{X+\frac{\Delta X}{2}}
-\theta\left(X-\frac{\Delta X}{2}
\right)
\sqrt{X-\frac{\Delta X}{2}}
\right],
\end{equation}
where $\theta(x)$ is the Heaviside unit step function. The density reaches its maximum 
at $X=\Delta X/2$ with $n_{e, \max}=n_0\sqrt{ m_e c/\Delta p_0}$. In the limit $X\gg \Delta X$ 
the electron density 
is inversely proportional to the square root of $X$, $n_e(X)\sim 1/\sqrt{X}$ as in the case 
corresponding to Eq. (\ref{eq-neabovecold}). 
In the laboratory frame of reference we have $n_{e, \max}=n_0\sqrt{ m_e c\gamma_{\rm ph}^3/\Delta p_0}$.
The contribution to the electron density from the cubic part of the distribution function is
proportional to the surface of the area bounded by the curves
$p_{\pm}(X)$ which are the roots of equation (\ref{eq-cubics}) given by  the expressions
\begin{equation} 
\label{eq:w-b-ppm-cub}
 p_{\pm}(X)=
 \frac{2\, 3^{1/3} r+ 2^{1/3}
 \left(\sqrt{81 X_{\pm}^2 -12 r^3}-9 X_{\pm}\right)^{2/3}}
	 {6^{2/3}\left(\sqrt{81 X_{\pm}^2 -12 r^3}-9 X_{\pm}\right)^{1/3}}
	 \end{equation}
with $X_{\pm}=X-X_c \pm \Delta X/2$, where $X$ is normalized on $mc/e E$ 
and $p$ measured in units of $m_e c$.
At $X\gg \Delta p$ the electron density is   proportional to $X^{-2/3}$, as in the case 
corresponding to Eq. (\ref{eq24-nbre}).
  
  We see an apparent similarity between the density distribution obtained with the computer simulations, 
 which is shown in Fig. \ref{fig11} f, and the density distribution given 
 by the simple model (Fig. \ref{fig12} b).

 When $r=0$ the cubic part of the distribution function  develops  a new breaking point  
 and for $r<0$  it  is no longer a single valued functions of $X$. 	
 At $r=0$  the contribution of the cubic part  results 
 in the electron density described by 
 \begin{equation} 
\label{eq:w-b-n_e-cub}
n_e(X) \sim 
\frac{n_0}{\Delta p_0}
\left[ 
\theta \left(X_{+}\right) \left(X_{+}\right)^{1/3}
+\theta \left(-X_{-}\right) \left(-X_{-}\right)^{1/3}
-
\theta \left(X_{-}\right) \left(X_{-}\right)^{1/3}
-\theta \left(-X_{+}\right) \left(-X_{+}\right)^{1/3}
\right].
\end{equation}
In the limit $\Delta p_0 \to 0$ the electron density profile for $X \to 0$ is given by  $n_e(X) \sim X^{-2/3}$
in accordance with  the theory of the wave breaking in a cold plasma (see Eq. (\ref{eq24-nbre})).

 \subsection{Energy scaling of laser accelerated electrons}
 
 Here we consider the LWFA acceleration in the above wave breaking regime when the wake field amplitude
is not limited by the value $E_{\rm AP}$ (\ref{eq-Emx}) and is related via Eq. (\ref{eq-Eabwbrmax}) 
for $\gamma_{\rm m}=1+a^2/2$ to the 
laser pulse amplitude as $E_{\max}=a$.
 The electron injected into the wakefield acceleration phase can acquire the energy  \cite{ESL1}
 \begin{equation} 
\label{eq:lwfa-energy}
{\cal E}=\frac{e \varphi_w}{1-\beta_{\rm ph}},
\end{equation}
  where the wakefield electrostatic potential is equal to $\varphi_w \approx 2 \pi n_0 e^2 r_w^2$ with $r_w$ 
  being
the wakewave transverse size, which is of the order of  the laser pulse waist equal to 
$r_w \approx (c/\omega_{\rm pe})\sqrt{2 a}$. The amplitude of the laser pulse is given by Eq. (\ref{eqSSB}). 
Using the relationship between the laser power and the amplitude (\ref{eqSSB}) and between the wake wave
 phase velocity and the plasma density, which can be written as $\gamma_{\rm ph}=\sqrt{n_{\rm cr} a/n_0}$ 
 (see Ref. \cite{EK}), we obtain for the accelerated electron energy 
  \begin{equation} 
\label{eq:lwfa-energypow}
{\cal E}\approx m_e c^2 \left(\frac{\cal P}{{\cal P}_c}\right)^{2/3}\left(\frac{n_{\rm cr}}{n_0}\right)^{1/3}.
\end{equation}

As we see, for given laser power the fast electron energy is proportional to $n_0^{-1/3}$, 
i.e. the lower plasma density, the higher the electron energy. 
The electron density cannot be lower than the density determining the relativistic 
self-focusing threshold 
(here we do not consider the laser wakefield excited inside a plasma waveguide, 
i.e. inside a plasma filled capillary), at which $n_{0,\min}=n_{\rm cr} {{\cal P}_c}/{\cal P}$   and 
$a\approx 1$, i.e. the wake plasma wave is in the weakly nonlinear regime 
as required for the laser based electron-positron collider \cite{LEPT}, i.e. for $a \ge 1$. 
As the result, we obtain the electron energy scaling under the optimal conditions
 \begin{equation} 
\label{eq:lwfa-energyopt}
{\cal E}\approx m_e c^2 \left(\frac{\cal P}{{\cal P}_c}\right),
\end{equation}
which for ${\cal P}=50$TW yields ${\cal E}=3$GeV, and for ${\cal P}=100$PW gives ${\cal E}=6$TeV.

The acceleration length according to Eq. (\ref{eq:lwfa-energy}), $l_{\rm acc}=2 r_w \gamma^2_{\rm ph}$,
in the optimal regime is given by 
\begin{equation} 
\label{eq:lwfa-lacc}
l_{\rm acc}\approx \frac{\lambda}{\pi} \left(\frac{\cal P}{{\cal P}_c}\right)^{3/2}.
\end{equation}
In the case of ${\cal P}=50$TW one-micron wavelength laser, we have $l_{\rm acc}\approx 5$cm.

We recall that in the limit of large laser amplitudes the  energy scaling  
of  the accelerated electrons  in Eq. (\ref{eq:lwfa-energyopt}) has  a different dependence 
on the laser plasma parameters as discussed in Ref. \cite{ESL1} 
and references quoted therein. 

\section{Discussions and Conclusions}

In the present paper, by extending an approach formulated in Ref. \cite{RCD} to the relativistic limit,
we investigated the wave breaking of relativistically strong Langmuir wave in thermal plasmas.
As is  well known, the wavebreak concept is meaningful only 
for systems which allow the hydrodynamics description because in kinetic systems with 
broad distribution functions there are always processes similar to wave breaking, such as
the Landau damping in linear 
and nonlinear 
 regimes.

In the study of high power laser matter interaction wavebreak-like processes attract great attention
in regimes where the wave amplitude is much larger than the distribution 
thermal spread in the momentum space, the most relevant questions being the maximal electric field, 
on the structure of the formed singularity and on the number of electrons involved in the wavebreaking.

Using the relativistic waterbag model we showed the typical structures of singularities occurring during the
wave breaking, we found the dependence of maximum electric field on the wave parameters, 
and discussed the behaviour
of nonlinear wave in collisionless plasmas. The approach based on the warm plasma fluid model \cite{ESL2} 
leads to the same scalings for the profile of the breaking waves. 

We found that in the above breaking
limit the electron distribution in the nonlinear wave takes a skewed form. 
Note the somewhat similar feature in 
breaking water waves, when a symmetric Stokes profile \cite{Stokes} evolves to a skewed wave 
(see Ref. \cite{Whitham}).

\medskip
\begin{acknowledgments}
We thank  A. G. Zhidkov for discussions. We acknowledge support of this work from the MEXT of Japan,
Grant-in-Aid for Scientific Research, 23740413 and Grant-in-Aid for Young Scientists 21740302 from MEXT. 
We appreciate support from the NSF under Grant No. PHY-0935197 and 
the Office of Science of the US DOE under Contract No. DE-AC02-05CH11231.

\end{acknowledgments}



\begin{thebibliography}{99}
\bibitem{GINZBURG} 
V. L. Ginzburg, {\it The Propagation of Electromagnetic Waves
in Plasmas} (Pergamon Press, Oxford, 1970); 
R. K. Dodd, J. C. Eilbeck, J. D. Gibbon, H. C. Norris, {\it Solitons and
Nonlinear Wave Equations} (Academic Press Inc., New York, 1984); 
W. L. Kruer, {\it Physics of Laser Plasma Interactions} (Addison-Wesley, Menlo
Park, CA, 1988);  
M. S. Longair, {\it High Energy Astrophysics} (Cambridge Univ. Press, Cambridge
1992).

\bibitem{LWFA} T. Tajima and J. M. Dawson,  \textit{Phys. Rev. Lett.} \textbf{34},
269 (1979). 

\bibitem{ESL1} E. Esarey, C. B. Schroeder, W. P. Leemans, \textit{Rev. Mod. Phys. }\textbf{81}, 1229 (2009).

\bibitem{PWFA} P. Chen, J. M. Dawson, R. W. Huff et al., \textit{Phys. Rev. Lett.} \textbf{54},
693 (1985); 
T. Katsouleas, \textit{Phys. Rev. A} \textbf{33},2056 (1986);
I. Blumenfeld, C. E. Clayton, F.-J. Decker et al., \textit{Nature} {\bf 445}, 741 (2007).

\bibitem{HOH} D. F. Gordon,  B. Hafizi, D. Kaganovich, A. Ting,  
\textit{Phys. Rev. Lett.} {\bf 101}, 045004 (2008);
U. Teubner and P. Gibbon, \textit{Rev. Mod. Phys.} \textbf{81}, 445 (2009); 
A. S. Pirozhkov, M. Kando, T. Zh. Esirkepov et al., \textit{Phys. Rev. Lett.} {\bf 108}, 135004 (2012).

\bibitem{MTB} G. Mourou, T. Tajima, S. V. Bulanov, \textit{Rev. Mod. Phys.} 
\textbf{78}, 309 (2006).

\bibitem{Part-II} 
S. V. Bulanov, T. Zh. Esirkepov, M. Kando, J. K. Koga, A. S. Pirozhkov, T. Nakamura, S. S. Bulanov, 
C. B. Schroeder, E. Esarey, F. Califano, and F. Pegoraro,
 {\it Phys. Plasmas} (2012) - submitted for publication; [arXiv e-print: 2012ArXiv1202.1907B].

\bibitem{AP} A.~I.~Akhiezer and R.~V.~Polovin, \textit{Sov. Phys. JETP} \textbf{30}, 915 (1956).

\bibitem{MaX} S. V. Bulanov, V. I. Kirsanov,  A. S. Sakharov, \textit{JETP Letters} \textbf{53}, 565 (1991).

\bibitem{RCD} R. C. Davidson, {\it Methods in nonlinear plasma theory} (Academic Press Inc., New York, 1972).

\bibitem{JMD} J. M. Dawson, \textit{Phys. Rev.} \textbf{113}, 383 (1959).

\bibitem{KM} T. Katsouleas and W. Mori, \textit{Phys. Rev. Lett.} \textbf{61},90 (1988).

\bibitem{Inj} S. V. Bulanov, I. N. Inovenkov, V. I. Kirsanov  
et al., \textit{Phys. Fluids B} \textbf{4}, 1935 (1992); C. A. Coverdale, C.
B. Darrow, C. D. Decker
et al., 
\textit{Phys. Rev. Lett.} \textbf{74}, 4659 (1995); 
A. Modena, A. Najmudin, E. Dangor
et al., \textit{Nature (London)} \textbf{377}, 606 (1995);
S. V. Bulanov, F. Pegoraro, A. M. Pukhov, A. S. Sakharov, {\it Phys. Rev. Lett.} {\bf 78},
4205 (1997);
D. Gordon, K. C. Tzeng, C. E. Clayton
et al., \textit{Phys. Rev. Lett.} \textbf{80}, 2133 (1998);
S. V. Bulanov, N. Naumova, F. Pegoraro, J. Sakai, \textit{Phys. Rev. E} \textbf{58}, R5257 (1998); 
H. Suk, N. Barov, J. B. Rosenzweig, E. Esarey, \textit{Phys. Rev. Lett.} \textbf{86}, 1011 (2001); 
A. Pukhov and J. Meyer-Ter-Vehn, {\it Appl. Phys. B} {\bf 74}, 355 (2002);
M. C. Thompson, J. B. Rosenzweig, H. Suk, \textit{Phys. Rev. ST Accel. Beams} \textbf{7}, 011301 (2004); 
P. Tomassini, M. Galimberti, A. Giulietti
et al.,
\textit{Laser Part. Beams} \textbf{22}, 423 (2004);
T. Ohkubo, A. G. Zhidkov, T. Hosokai et al. \textit{Phys. Plasmas} \textbf{13}, 033110 (2006);
M. Kando, Y. Fukuda, H. Kotaki, et al., {\it JETP}, {\bf 105}, 916 (2007);
C. G. R. Geddes, K. Nakamura, G. R. Plateau
et al., \textit{Phys. Rev. Lett.} \textbf{100}, 215004 (2008);
A. V. Brantov, T. Zh. Esirkepov, M. Kando
et al., \textit{Phys. Plasmas} \textbf{15}, 073111 (2008);
J. Faure, C. Rechatin, O. Lundh
et al., \textit{Phys. Plasmas} \textbf{17}, 083107 (2010); 
K. Schmid, A. Buck, C. M. S. Sears, et al. {\it Phys. Rev. ST Accel. Beams} \textbf{13}, 091301 (2010);
Y.-C. Ho, T.-S. Hung, C.-P. Yen
et al., \textit{Phys. Plasmas} \textbf{18}, 063102 (2011); 
A. J. Gonsalves, K. Nakamura, C. Lin
et al., \textit{Nature Phys.} \textbf{7}, 862 (2011);
Y. Y. Ma, S. Kawata, T. P. Yu, et al., \textit{Phys. Rev. E} \textbf{85}, 046403 (2012).

\bibitem{ESL2} 
C. B. Schroeder, E. Esarey, B. A. Shadwick, \textit{Phys. Rev. E} \textbf{72,} 055401 (2005); 
C. B. Schroeder, E. Esarey, B. A. Shadwick, W. P. Leemans, \textit{Phys. Plasmas }\textbf{13,} 033103 (2006);
C. B. Schroeder, E. Esarey, B. A. Shadwick, \textit{Phys. Plasmas } \textbf{14}, 084701 (2007);  
C. B. Schroeder and E. Esarey, {\it Phys. Rev. E} \textbf{81}, 056403 (2010).
 

\bibitem{TR} R. M. G. M. Trines and P. A. Norreys, \textit{Phys. Plasmas } \textbf{13}, 123102 (2006); 
R. M. G. M. Trines and P. A. Norreys, \textit{Phys. Plasmas }\textbf{14}, 084702 (2007); 
R. M. G. M. Trines, \textit{Phys. Rew. E }\textbf{79,} 056406 (2009);
R. M. G. M. Trines, R. Bingham, Z. Najmudin et al. {\it New Jornal of Physics} {\bf 12}, 045027 (2010);
Z. M. Sheng and J. Meyer-ter-Vehn,  \textit{Phys. Plasmas }\textbf{4}, 493 (1997).

\bibitem{COF} T. P. Coffey, \textit{Phys. Fluids} \textbf{14,} 1402 (1971); 
T. Coffey, \textit{Phys. Plasmas} \textbf{17,} 052303 (2010); 

\bibitem{BurNob} D. A. Burton and A. Noble, \textit{J. Phys. A: Math. Theor.} \textbf{43}, 075502 (2010).

\bibitem{SMF} A. A. Solodov, V. M. Malkin, N. J. Fisch, {\it Phys. Plasmas} {\bf 13}, 093102 (2006).

\bibitem{PAN} A. V. Panchenko, T. Zh. Esirkepov, A. S. Pirozhkov et al., 
\textit{Phys. Rev. E} \textbf{78}, 056402 (2008).

\bibitem{BRMNN} B. Riemann, \textit{Abhandlungen der K\"{o}niglichen
Gesellschaft der Wissenschaften zu G\"{o}ttingen}, \textbf{8}, 43 (1860).

\bibitem{LLHd} L. D. Landau and E. M. Lifshitz, {\it Fluid Mechanics} (Butterworth and Heinemann, Oxford, 1987).

\bibitem{Stokes} G. G. Stokes, \textit{Trans. Cambridge Philos. Soc.} {\bf 8}, 441 (1847);
J. Wilkening, \textit{Phys. Rev. Lett.} {\bf 107}, 184501 (2011).

\bibitem{Whitham} G. B. Whitham, {\it Linear and Nonlinear Waves} (Wiley-Interscience, New York, 1974).

\bibitem{peakon} R. Camassa and D. D. Holm, \textit{Phys. Rev. Lett.} {\bf 71}, 1661 (1993).

\bibitem{KCP} L. V. Keldysh, {\it Sov. Phys. JETP} {\bf 20}, 1307 (1965);
V. S. Popov, {\it Phys. Usp.} {\bf 47}, 855 (2004).

\bibitem{Koga2011} J. K. Koga et al., in preparation.


\bibitem{PORPEG}  F. Pegoraro and F. Porcelli, {\it Phys. Fluids}, {\bf  27}, 1665 (1984).

\bibitem{buti}    B. Buti {\it  Phys. Fluids}, {\bf  5}, 1 (1962).

\bibitem{NPIC} T. Nakamura, M. Tampo, R. Kodama et al.,
{\it Phys. Plasmas} {\bf 17}, 113107 (2010).

\bibitem{EK} A. Zhidkov, J. Koga, K. Kinoshita, M. Uesaka, 
{\it Phys. Rev. E} {\bf 69}, 035401(R) (2004).

\bibitem{RSF} G. A. Askar'yan, {\it Sov. Phys. JETP} {\bf 15}, 1088 (1962); 
A. G. Litvak, {\it Sov. Phys. JETP} {\bf 30}, 344 (1969); 
C. E. Max, J. Arons, A. B. Langdon, {\it Phys. Rev. Lett.} {\bf 33}, 209 (1974); 
P. Sprangle, C. M. Tang, E. Esarey, {\it IEEE Trans. Plasma Sci.} {\bf 15}, 145 (1987); 
G. Z. Sun, E. Ott, Y. C. Lee, P. Guzdar, {\it Phys. Fluids} {\bf 30}, 526 (1987); 
A. B. Borisov, A. V. Borovskiy, V. V. Korobkin et al., 
{\it Phys. Rev. Lett.} {\bf 65}, 1753 (1990); P. Monot, T. Auguste, P. Gibbon et al., 
{\it Phys. Rev. Lett.} {\bf 74}, 2953 (1995).
\bibitem{FRCA} A. Mangeney, F. Califano, C. Cavazzoni, P. Travnicek, 
{\it J. Comp. Physics} {\bf 179}, 495 (2002). 

\bibitem{SSB} S. S. Bulanov, V. Yu. Bychenkov, V. Chvykov et al., 
{\it Phys. Plasmas} {\bf 17}, 043105 (2010).

\bibitem{AS} M. Abramowitz and I. A. Stegun, {\it Handbook of Mathematical
Functions with Formulas, Graphs, and Mathematical Tables}
 (Dover, New York, 1964).

\bibitem{LEPT} W. Leemans and E. Esarey, {\it Physics Today} {\bf 62}, 44 (2009).

\end{thebibliography}
\end{document}